\documentclass[11pt,a4paper]{article}
\usepackage{jheppub,amsmath,  amssymb,slashed,url,bm,textgreek,upgreek}
\usepackage{graphicx}
\usepackage{epstopdf}
\def\t{{ \sf t}} 
\def\fF{{\sf F}}
\def\tt{{\mathfrak t}}

\def\a{{\sf a}}
\def\tach{{\mathrm{tach}}}
\def\gauge{{\mathrm{gauge}}}
\def\tr{{\mathrm{tr}}}
\def\veps{\varepsilon}

\def\Tr{{\mathrm{Tr}}}

\def\SO{{\mathrm{SO}}}
\def\OO{{\mathrm{O}}}
\def\Spin{{\mathrm{Spin}}}

\def\UU{{\mathrm U}}

\def\be{\begin{equation}}
\def\ee{\end{equation}}

\def\h{\widehat}

\def\A{{\mathcal A}}
\def\op{{\mathrm{op}}}
\def\d{{\mathrm d}}

\def\b{{\sf b}}
\def\R{{\mathbb R}}
\def\C{{\mathbb C}}

\def\[{\bigl [}

\def\]{\bigr ]}

\def\tT{{\sf T}}

\def\tr{{\mathrm{tr}}}
\def\Z{{\mathbb Z}}

\def\t{\widetilde }
\def\h{\widehat}

\def\i{{\mathrm i}}

\def\B{{\mathcal B}}

\def\la{\langle}
\def\ra{\rangle}
\def\Pf{{\mathrm{Pf}}}

\def\H{{\mathcal H}}

\def\bar{\overline}

\def\oO{{\mathrm O}}

\def\rR{{\sf R}}
\def\SU{{\mathrm{SU}}}
\def\Sp{{\mathrm{Sp}}}

\font\teneurm=eurm10 \font\seveneurm=eurm7  \font\fiveeurm=eurm5
\newfam\eurmfam
\textfont\eurmfam=\teneurm \scriptfont\eurmfam=\seveneurm
\scriptscriptfont\eurmfam=\fiveeurm

\font\teneusm=eusm10 \font\seveneusm=eusm7 \font\fiveeusm=eusm5
\newfam\eusmfam
\textfont\eusmfam=\teneusm \scriptfont\eusmfam=\seveneusm
\scriptscriptfont\eusmfam=\fiveeusm

\font\tencmmib=cmmib10 \skewchar\tencmmib='177
\font\sevencmmib=cmmib7 \skewchar\sevencmmib='177
\font\fivecmmib=cmmib5 \skewchar\fivecmmib='177
\newfam\cmmibfam
\textfont\cmmibfam=\tencmmib \scriptfont\cmmibfam=\sevencmmib
\scriptscriptfont\cmmibfam=\fivecmmib

\title{Anomalies and Non-Supersymmetric D-Branes}

 \author{Edward Witten}
\affiliation{School of Natural Sciences, Institute for Advanced Study,\\ 1 Einstein Drive, Princeton, NJ 08540 USA}
\abstract{We revisit some aspects of D-brane theory from the point of view of anomalies.  When the boundary condition on a worldsheet boson is flipped from Neumann
to Dirichlet, worldsheet supersymmetry requires also reversing the sign of the boundary condition of the corresponding worldsheet fermion.   This induces an anomaly
which is a mod 2 anomaly in Type II superstring theory and a mod 8 anomaly in Type I superstring theory.   The same anomaly also receives contributions from a sign
in the sum over bulk spin structures (in Type IIA superstring theory), Chan-Paton factors of symplectic type (in Type I superstring theory), 
and Majorana fermions that propagate only on the worldsheet boundary.    The need to cancel the anomaly accounts for many properties of supersymmetric and
especially nonsupersymmetric D-branes in Type I and Type II superstring theory.
}

\begin{document}\maketitle

\section{Introduction}\label{intro} 

 The most important D-branes in string theory are the supersymmetric ones \cite{DLP}.   In particular,
 Type IIA (IIB) superstring theory has supersymmetric Dp-branes of even (odd) $p$, which play a key role in 
 the string duality picture \cite{Pol,Johnson}.
 
 It is also possible to construct nonsupersymmetric D-branes, that is, D-branes that are not invariant
 under any supersymmetries \cite{Sen1,Sen2,Sen3}.   In Type II superstring theory, there are nonsupersymmetric Dp-branes with $p$ odd for Type IIA and even for Type IIB.   
 They are unstable, which limits their relevance in certain applications.  There are also
 nonsupersymmetric D-branes in Type I superstring theory, some of which are stable.  In particular, Type I superstring theory has stable
 D0- and D$(-1)$-branes, which are important in understanding the duality between Type I superstring theory and the heterotic string \cite{Sen1,Sen2,Sen3}.   This work
 involved considerations of tachyon condensation along branes to generate new branes of lower dimension, but also some explicit conformal field theory constructions.  
 Type I superstring theory also has a rich spectrum of unstable Dp-branes for various $p$, which were described and analyzed in \cite{Bergman,ATS}, mostly from the point of
 view of tachyon condensation and K-theory.

In the present article, we will revisit the construction and properties of nonsupersymmetric D-branes from the point of view of fermion anomalies.
As we will discuss, a two-dimensional worldsheet theory that describes a Type I or Type II superstring, although anomaly-free in bulk, is potentially anomalous
when formulated on a worldsheet with boundary.   Here we will only explore the simplest aspect of this question, which depends only on the dimension of the D-brane 
world-volume.\footnote{A more detailed study of the anomaly in string theory with D-branes leads to understanding that in Type II
superstring theory, the ``$\UU(1)$ gauge bundle'' on a D-brane worldvolume is really a ${\mathrm{Spin}}_c$ structure \cite{FW}, a statement that is an aspect of the relation
between D-branes and K-theory.  We will not revisit those issues here.} 
  In the case of Type II superstring theory, the worldsheet is oriented and the anomaly is $\Z_2$-valued.    This anomaly has applications in various areas of physics;
  for an application in condensed matter physics, see \cite{Kitaev}.
The restriction for $p$ to be even or odd in a supersymmetric Dp-brane of Type IIA or IIB  can be regarded as a consequence of anomaly cancellation.
If one wishes to reverse the parity of $p$, which by itself would make the worldsheet theory anomalous, one can cancel the anomaly by adding a single Majorana fermion
mode $\lambda$  that propagates only on the worldsheet boundary.    This method of anomaly cancellation leads to a nonsupersymmetric Dp-brane, with reversed values
of $p$, that possesses some unusual properties that were first identified in \cite{Sen1,Sen2,Sen3}, where these branes were constructed in a quite different way: the tension of this
brane exceeds that of a supersymmetric D-brane of the same $p$
(which exists for the opposite Type IIA or IIB theory) by a factor of $\sqrt 2$, and in the conventional sense (that is, if one does not explicitly take into account the $\lambda$ mode)
it has no GSO projection.   

For Type I superstring theory, instead, the anomaly is $\Z_8$-valued.   This fact is related to the properties of real Clifford algebras \cite{ABS} and has had important applications
in condensed matter physics \cite{FK}.   As in the Type II case, some  properties of supersymmetric Type I Dp-branes -- notably the fact that the Chan-Paton symmetry groups of
D1-branes and D9-branes are orthogonal, while D5-branes have symplectic symmetry groups -- can be naturally understood in terms of anomaly cancellation.
For other values of $p$, anomaly cancellation leads to nonsupersymmetric Dp-branes, including the D0-branes that were described in \cite{Sen1,Sen2,Sen3} and shown to have
important applications to heterotic-Type I duality, as well as unstable ones that were originally described and studied in \cite{Bergman,ATS}.

In this article, we are mostly revisiting results that have already been obtained in other ways.   Apart from papers already cited,
some of the relations to previous work are as follows. 
  In \cite{KL}, worldsheet theories that govern supersymmetric and nonsupersymmetric
Dp-branes were related to each other via boundary renormalization group flows.   Anomalies were not invoked explicitly, but because the relevant anomalies are invariant
under renormalization group flows, the construction makes it manifest that certain boundary conditions are anomaly-free if and only if other ones are.  (Tachyon
condensation, an important tool in \cite{Sen1,Sen2,Sen3,Bergman,ATS}, has the same property of never generating anomalies from an anomaly-free starting point.)
Majorana modes that propagate only on the boundary are natural in
the setup of \cite{KL} and a number of things described in this article have natural explanations in that setup.  
Anomalies and their relation to boundary fermions were considered explicitly and in detail in  \cite{KPT,KPT2}, in exploring questions
very closely related to the considerations of the present article.   Those authors explored many of the matters that we will analyze here, from a different point of view,
and analyzed  a number of issues that we will not consider, including Type 0 strings and orientifolds.   On some questions, in the present article we are simply giving more detailed
and possibly more elementary explanations than the ones given already in \cite{KPT,KPT2}.

Nonsupersymmetric D-branes have also been studied via supergravity, for example in \cite{SuperA,SuperB,SuperC}.

In section \ref{TypeII}, we discuss the $\Z_2$-valued anomaly of Type II superstring theory and its relation to nonsupersymmetric D-branes.     
Much of the content of this section was presented in lectures at conferences in honor of Martin Rocek and
in memory of Shoucheng Zhang  \cite{Lecture1,Lecture2}.    The $\Z_8$-valued anomaly
for Type I and its implications for D-branes is the topic of section \ref{TypeI}.

\section{Branes In Type II Superstring Theory}\label{TypeII}

\subsection{Three Manifestations Of A $\Z_2$-Valued Anomaly}\label{threety}

Three ingredients that all carry the same mod 2 anomaly will be important in our discussion of Type II D-branes.
Anomaly-free branes can be constructed by combining the three ingredients in any combination such that the anomaly cancels.\footnote{A fourth ingredient that also carries
the same anomaly was important in \cite{DW}, where some of the following background material was also presented.
 In section \ref{TypeI} we will encounter yet another ingredient that contributes to the Type I version of this anomaly, namely
Chan-Paton factors of symplectic type.}   The most elementary of the three is a Majorana fermion $\lambda$ in one spacetime dimension.   The action is
\be\label{zelbo} I=\int\d \tau \frac{\i}{2}\lambda\frac{ \d\lambda}{\d \tau}.  \ee
The Hamiltonian of this theory vanishes.
The theory has one important global symmetry, namely the symmetry $(-1)^\fF$ that acts by $\lambda\to -\lambda$.   The theory also has a time-reversal symmetry, which is not
important for Type II superstring theory but will be important when we discuss Type I in section \ref{TypeI}.   In our application, the 1-manifold on which this theory is formulated
will be the boundary of a string worldsheet.

Let us discuss this theory in Euclidean signature on a compact 1-manifold, which inevitably is a circle $S$.   $S$ comes with two possible spin structures, as $\lambda$ may be
periodic or antiperiodic in going around the circle.  The two possibilities correspond respectively to the Ramond  and Neveu-Schwarz(NS)  spin structures in string theory.   In the
case of a Ramond spin structure, $\lambda$ has a single zero-mode -- a constant mode on the circle.   This actually immediately implies that the theory is anomalous;
the path integral measure is odd under $(-1)^\fF$.   On a circle of circumference $2\pi$, we can expand $\lambda=\frac{1}{\sqrt{2\pi}}\sum_{n\in \Z} e^{-\i n \tau}\lambda_n$.
The action is then $I=\sum_{n=1}^\infty n\lambda_{-n}\lambda_n$.   The nonzero modes are paired; each pair $\lambda_n,\lambda_{-n}$ has a measure $\d\lambda_n \d\lambda_{-n}$
that is invariant under $(-1)^\fF$.  We are left with one unpaired fermion mode $\lambda_0$, so overall the measure is odd.  Another way to explain this is as follows.   If we pick
an integration measure for the path integral, then the partition function of the $\lambda$ field on a circle with Ramond spin structure will certainly vanish, because of the zero-mode.
But an insertion of $\lambda$ will lift the zero-mode and give $\la\lambda\ra\not=0$.   Since $\lambda$ is odd under $(-1)^\fF$, the fact that the path integral computes a nonzero
value of $\la\lambda\ra$ means that the path integral measure is odd under $(-1)^\fF$; this is the anomaly.

Clearly this is a mod 2 anomaly, as it only involves the sign of the path integral measure.   If instead of a single Majorana mode $\lambda$, we consider $k$ such modes
$\lambda_1,\cdots,\lambda_k$, then in the Ramond sector there are $k$ zero-modes, and there is an anomaly in $(-1)^\fF$ if and only if $k$ is odd.

In the NS sector, there is no anomaly, but nonetheless by studying the path integral in the NS sector, one can see that the theory of a single Majorana fermion is problematical.
There is only a single canonical variable $\lambda$, and the canonical anticommutation relations reduce to $\lambda^2=1$.   In an irreducible representation of this algebra,
$\lambda$ is a $c$-number 1 or $-1$, and the Hilbert space is 1-dimensional.    However, in such an irreducible representation, the symmetry operator $(-1)^\fF$ cannot
be implemented; it exchanges the two irreducible representations of the anticommutation relations with $\lambda=1$ and $\lambda=-1$.   In order to implement the $(-1)^\fF$ operator,
we need to include both representations, so we need a two-dimensional Hilbert space.

What value of the Hilbert space dimension does the path integral favor?   The answer is simply given by the $\lambda$ path integral in the NS sector. 
In a generic quantum theory with Hamiltonian $H$ and quantum Hilbert space $\H$, the path integral on a circle of circumference $\beta$ with NS spin structure 
computes $\Tr_\H\,e^{-\beta H}$.
In the present example, $H=0$ and this reduces to the dimension of $\H$.
   However, the NS sector path integral actually produces the answer $\sqrt 2$, a nonsensical value
for the Hilbert space dimension.   One way to obtain the answer $\sqrt 2$ is to use zeta function regularization.   The eigenvalues of the fermion kinetic operator $D=\i\frac{\d}{\d \tau}$
are $n+1/2$, $n\in\Z$.  The fermion path integral is the Pfaffian of $D$,  formally  $\Pf(D)=\prod_{n=0}^\infty (n+1/2)$.   
To regularize this, we introduce the zeta function $\xi(s)=\sum_{n=0}^\infty(n+1/2)^{-s}$
and define the Pfaffian as $\Pf(D)=\exp(-\xi'(0))$.  Since $\xi(s)=(2^s-1)\zeta(s)$, where $\zeta(s)=\sum_{n=1}^\infty n^{-s}$ is the Riemann zeta function, and $\zeta(0)=-1/2$,
this procedure leads to $\Pf(D)=\sqrt 2$.

A less computational way to get the same answer is to consider first the anomaly-free theory with two Majorana fermions.   The operator algebra is now $\{\lambda_i,\lambda_j\}=2\delta_{ij}$, $i,j=1,2$.  This algebra has a straightforward realization in a two-dimensional Hilbert space, and the operator $(-1)^\fF$ can be realized in this space as $(-1)^\fF=\i\lambda_1\lambda_2$.
So the NS sector path integral for two Majorana fermions gives the answer 2, and therefore for one Majorana fermion the answer must be $\sqrt 2$.   The numerical value shows again
that the theory of a single Majorana fermion is problematical, but we will see how it plays a role in understanding nonsupersymmetric D-branes.

For our second example of a mod 2 anomaly, we start in 2 dimensions. Consider a two-component Majorana fermion $\psi$  in two dimensions. 
To write a Dirac operator, we need a pair of gamma matrices $\gamma_i,\,i=1,2$. We can pick them to be real $2 \times  2$ matrices (for example, the Pauli matrices
$\sigma_1$ and $\sigma_3$). Then the Dirac operator 
$\slashed{D} =\sum_i \gamma^i D_i$ is a real, skew-Hermitian operator. The path integral of a single Majorana fermion is then formally $\Pf(\slashed{D})$, the 
Pfaffian of the Dirac operator $\slashed{D}$.   In general, fermion anomalies affect only the phase of the path integral, since the absolute value of the path integral can
always be defined (for example via Pauli-Villars or $\zeta$-function regularization) in a way that respects all symmetries.
In the present example, 
the fermion Pfaffian is naturally real, since the operator itself
is real, so the only possible anomaly would involve the sign of the Pfaffian.

To get a bulk theory that definitely has no anomaly, we can consider a pair of identical real Majorana fermions. The path integral is then
the square of the Pfaffian; it is positive, and thus completely anomaly-free.   

   \begin{figure}
 \begin{center}
   \includegraphics[width=1.5in]{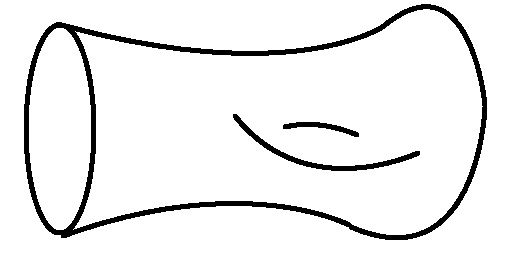}
 \end{center}
\caption{\footnotesize  A Riemann surface with boundary.  \label{one}}
\end{figure} 
 
 Now let us consider this theory on a Riemann surface $\Sigma$ with boundary (fig. \ref{one}).
There are two natural boundary conditions that preserve the skew-symmetry of the Dirac operator and are usually used in D-brane physics, namely
\be\label{twob}\bigl.(\gamma\cdot n )\psi\bigr| =\bigl.\pm\psi\bigr|,\ee
where $n$ is the normal vector to the boundary.  The two boundary conditions are equally good; they are exchanged by a chiral rotation
$\psi\to \bar\gamma\psi$, $\bar\gamma=\gamma_1\gamma_2$.

If we want a theory on a two-manifold with boundary
that is definitely anomaly-free, we can take a pair of Majorana fermions each with the same boundary condition on all boundary components of $\Sigma$.   
Then the path integral is still $(\Pf(\slashed{D}))^2$ and it is still positive, and hence anomaly-free.

\begin{figure}
 \begin{center}
   \includegraphics[width=2.5in]{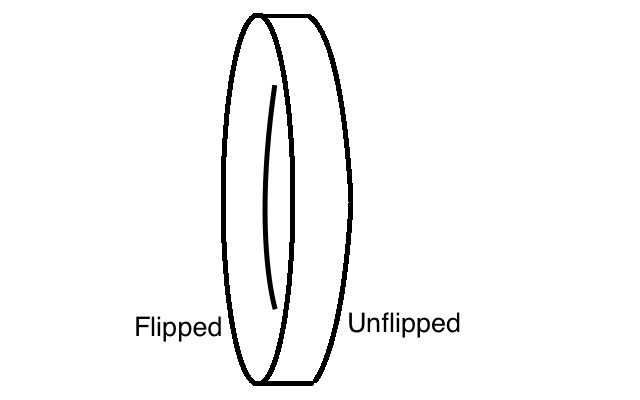}
 \end{center}
\caption{\footnotesize    An annulus $S^1\times I$.  If   two Majorana fermions satisfy the same boundary condition on one boundary of the
annulus  but opposite boundary conditions on the other boundary, then precisely one of the two fermions has a zero-mode along $I$.   The theory then reduces
along $S^1$ to the anomalous theory of a single Majorana fermion in one spacetime dimension.    \label{two}}
\end{figure}

Suppose, however, that we flip the boundary condition for one of the two Majorana fermions on one (or more) of the boundary components. The theory is then
anomalous, with in a sense the same anomaly that we found in the purely 1-dimensional theory of a single Majorana fermion.  
One way to see this is to observe that the theory with the flipped boundary condition on one boundary can be compactified
to the anomalous 1d theory that we had before.   To do this, we take $\Sigma$ to be an annulus $S^1\times I$ (fig. \ref{two}), where $S^1$ is a circle and $I$ is an interval.
If the two Majorana fermions have the same boundary conditions on both boundaries of the annulus, then as just explained the theory is completely anomaly-free. 
Suppose, however, that the two Majorana fermions satisfy the same boundary condition on one boundary of the annulus, and satisfy opposite boundary conditions on the other
boundary.   Then one of the two fermions, but not the other, has a zero mode in the $I$ direction, and the theory therefore reduces along $S^1$ to the anomalous one-dimensional
theory that we investigated already.

We will explain another way to see that the ``flipped'' boundary condition produces an anomaly,  related to developments in condensed matter physics \cite{Kitaev}.
  Add a fermion bare mass so that the Dirac equation becomes 
\be\label{Deqn} 0=(\slashed{D}+m)\psi =(\gamma^1 D_1 +\gamma^2 D_2 +m)\psi.\ee
   Consider this equation on the half-space $x^1\geq 0$ in the $x^1-x^2$ plane,  and look for a zero-mode that is localized along the boundary. 
For a mode independent of $x^2$, and on a flat half-space, the equation reduces to
\be\label{loceq} \frac{\partial \psi}{\partial x} =-m\gamma_1\psi,\ee
where we set $x=x^1$.     So
\be\label{zold} \psi(x)=\exp(-m\gamma_1 x)\psi(0). \ee
 If the boundary condition is
\be\label{beqn} \gamma_1\psi(0)=\varepsilon\psi(0), ~~~\varepsilon=\pm 1,\ee
then
\be\label{neqn} \psi(x)=\exp(-m\varepsilon x)\psi(0).\ee
This is localized along the boundary $x=0$ if and only if 
$m\varepsilon>0.$  
For $m\varepsilon>0$, we get at low energies a single boundary-localized Majorana fermion mode.    As we learned at the outset, such a mode carries an anomaly.

Flipping the boundary condition for a single two-dimensional Majorana fermion will therefore either add or remove one boundary-localized Majorana fermion mode, depending on the
sign of $m$.   The bulk theory with two fermions of the same nonzero mass remains trivial at long distances: the only macroscopic effect of flipping the boundary condition is to add or remove one boundary-localized  Majorana mode.
  Therefore, flipping the boundary condition for one fermion shifts the boundary anomaly by one unit.

In the string theory application, the parameter $m$ is artificial; we really want $m=0$.  However, the conclusion that flipping the boundary condition shifts the boundary anomaly
by one unit does not depend on $m$, since in general anomalies do not depend on masses.     For our purposes, turning on $m$ is just one way to 
make the dependence of the anomaly on the boundary condition obvious.

 \begin{figure}
 \begin{center}
   \includegraphics[width=1.6in]{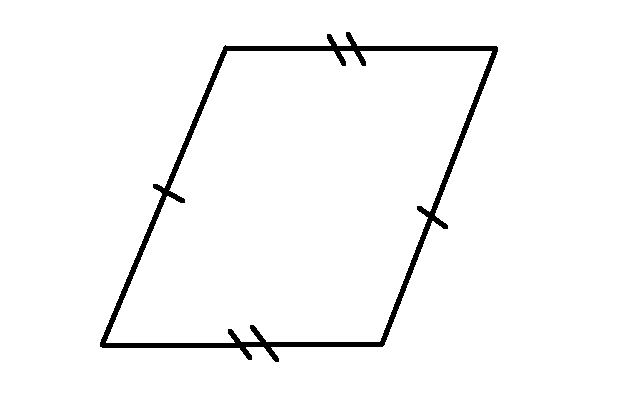}
 \end{center}
\caption{\footnotesize   Opposite sides are identified to make a  Riemann surface of genus 1.  Spin structures are conveniently labeled as $\pm \pm$, where the first (second) sign indicates whether a fermion
is periodic or antiperiodic in the horizontal (vertical) direction.    \label{three}}
\end{figure} 
We will need to understand one more way that the same anomaly can arise.     First let us discuss free fermions on an oriented  Riemann surface $\Sigma$ without boundary.     An important fact is
 that there are two types of spin structure, ``even'' and ``odd.''     To make this distinction, we consider a positive chirality fermion $\chi$ on $\Sigma$.  
 The Dirac action
 \be\label{dirac}\frac{1}{2}\int_\Sigma \d^2x \sqrt g \bar\chi \slashed{D} \chi\ee
 is antisymmetric, by fermi statistics.   
 The canonical form of an antisymmetric matrix is \be\label{canonical}\begin{pmatrix} 0 & -a & &&&&&& \cr a&0&&&&&&&\cr &&0&-b&&&&&\cr  &&b&0&&&\cr &&&&~~\ddots~~&&&&\cr &&&&&0& &&\cr &&&&&&0&&\cr &&&&&&&&~\ddots  \end{pmatrix},\ee
with nonzero modes that come in pairs and zero modes that are not necessarily paired.      The number of zero modes can change only when one of the ``skew eigenvalues''
$a,b,\cdots$ becomes zero or nonzero, and when this happens, the number of zero-modes jumps by 2.      So the number of zero-modes mod 2 is a topological invariant, called the mod
2 index, in this case the mod 2 index of the chiral Dirac operator.  We will denote it as $\zeta$.

As an example, consider a Riemann surface of genus 1 (fig. \ref{three}).
 There are four spin structures, usually labeled $\pm \pm$.     The $++$ spin structure has a single positive chirality zero-mode  (the ``constant'' mode of $\chi$) and the other spin
 structures have none.   So the $++$ spin structure is odd, with $\zeta=1$, and the others are even, with $\zeta=0$. 
 
 On a Riemann surface without boundary, a spin structure thus has the $\Z_2$-valued topological  invariant $(-1)^\zeta$.    
  This means that a theory with fermions on a closed oriented
Riemann surface can
have a discrete theta-angle, a factor $(-1)^\zeta$ that we include in the path integral measure. 
  However, $(-1)^\zeta$ cannot be defined, as a topological
 invariant, on a Riemann surface with boundary 
because there is no  satisfactory boundary condition on the chiral Dirac operator that can be used for that purpose.   The standard fermion boundary conditions of eqn. (\ref{twob})
are not suitable as they mix the two chiralities; thus, these boundary conditions do not make sense for a fermion that is supposed to have just one chirality, as assumed in the definition
of the mod 2 index.   One can actually show that the chiral Dirac operator does not admit any local, elliptic boundary condition.\footnote{It does admit a nonlocal
boundary condition that was introduced by  Atiyah, Patodi, and Singer in relation to the $\eta$-invariant.   However, this boundary condition does not enable a definition of
the mod 2 index as a topological invariant.}

 It turns out that, trying to define $(-1)^\zeta$ on a Riemann surface
with boundary, we run into the same mod 2 anomaly as before.  Before explaining this, we note the following.    It does not matter if we use positive chirality fermions or negative chirality fermions in defining $\zeta$;  the mod 2 index
is the same whether for fermions of positive or negative chirality.   This statement is true regardless of the choice of representation of the gamma matrices,
but it is most obvious if we choose 
 the gamma matrices to be real.   Then the Dirac equation is real and has a symmetry of complex conjugation.
The operator that has eigenvalue 1 or $-1$ for fermions of positive or negative chirality is
\be\label{chiral} \widehat\gamma=\i \bar\gamma=\i\gamma_1\gamma_2. \ee
This operator is imaginary, if the $\gamma_i$ are real,
 so its sign is reversed by complex conjugation.  Hence complex conjugation exchanges positive chirality fermion zero-modes with negative chirality ones, and the mod 2 index is the same for either chirality.

Now let us consider a Majorana fermion $\psi$ on a closed, oriented two-manifold $\Sigma$.     The action is
\be\label{undu} \int\d^2x\sqrt g\frac{1}{2}\bar\psi\slashed{D}\psi. \ee
To define this theory requires a choice of spin structure on $\Sigma$, so as already noted this theory has a discrete theta-angle: we can choose to modify the theory
by multiplying the path integral, for any chosen spin structure, by $(-1)^\zeta$.   
Classically there is a discrete chiral symmetry
\be\label{discchi}\psi\to\widehat\gamma\psi\ee  that multiplies fermions of positive or negative chirality by 1 or $-1$, respectively.
Quantum mechanically, this symmetry has an anomaly.     The symmetry assigns the value $-1$ to every negative chirality fermion zero-mode, so if there are $s$ such modes, then the
zero-mode measure transforms as $(-1)^s=(-1)^\zeta$ (where $\zeta$ is just the mod 2 reduction of $s$).    In other words, the chiral rotation flips the value of the discrete $\theta$-angle,
adding or removing a factor of $(-1)^\zeta$.

Suppose instead that the fermion has a bare mass $m$:
\be\label{ndu} \int\d^2x\sqrt g\bar\psi(\slashed{D}+m)\psi. \ee
The chiral rotation will now flip the sign of $m$ and also flip the value of the discrete theta-angle.

To make sure that there is no ambiguity in the sign of the fermion path integral, we can start with a {\it pair}
of Majorana fermions, with the same sign of $m$:
\be\label{helm} \int \d^2 x\sqrt g\sum_{i=1,2}\bar\psi_i(\slashed{D}+m)\psi_i.\ee
  If the fermions have the same mass $m$, the path integral is manifestly positive and the theory is completely trivial at long distances (modulo nonuniversal terms that
  can be eliminated by local counterterms).      So
integrating out a pair of massive fermions of the same mass gives a trivial theory at long distances.    If instead we make a chiral rotation
so that one of the two fermions has negative mass, the path integral will be
\be\label{zelm}(-1)^\zeta \Pf(\slashed{D}+m)\Pf(\slashed{D}-m),\ee
which must be equivalent.     This is trivial at long distances since it equals the partition function of the theory with both masses of the same sign, which we argued
to be trivial at long distances.  Instead of saying that the product in (\ref{zelm}) is trivial at long distances, an equivalent statement is that
  on a two-manifold without boundary, at long distances 
\be\label{telm}  \Pf(\slashed{D}+m)\Pf(\slashed{D}-m)=(-1)^\zeta ,\ee modulo local counterterms.

These considerations are relevant to string theory -- even though in string theory the worldsheet fermions are massless --  because in general anomalies do not depend on masses.
   Since a pair of fermions with the same masses is manifestly anomaly-free on a manifold without boundary, 
   a pair of fermions with opposite sign masses is also anomaly-free.   
So eqn. (\ref{telm}) implies that $(-1)^\zeta$ is anomaly-free on a manifold without boundary, as we indeed know to be true.   

Now let us consider a pair of fermions with opposite sign masses, but on a Riemann surface with boundary.
We give the two fermions the same boundary condition to make sure there is  still no anomaly.     In bulk they will generate
the same $(-1)^\zeta$ that was just described.     But on the boundary, because they have the same boundary condition
but opposite sign masses, they have the opposite sign of $m\varepsilon$ (in the notation of eqn. (\ref{neqn})), so one fermion field has a boundary-localized mode
and one does not.  Hence the boundary theory is anomalous.

The conclusion is that 
although $(-1)^\zeta$ is well-defined on a Riemann surface without boundary, on a Riemann surface with boundary
it has the same anomaly as a single real fermion mode $\lambda$ that propagates only on the boundary, 
with the action (\ref{zelbo}) with which we began.    In sum, then, we have found three things that have the same boundary-localized anomaly:
a Majorana fermion that propagates on the boundary; a pair of fermions with opposite boundary conditions; and a bulk factor $(-1)^\zeta$ in the path integral.

\subsection{Type II Superstring Theory}\label{super}

We are now ready to consider Type IIB and Type IIA superstring theory.     For Type II, on a Riemann surface without boundary, left- and right-moving (or negative and positive chirality)
worldsheet fermions see different spin structures, in general, and this is extremely important in constructing a tachyon-free theory with spacetime supersymmetry.   
For Type IIB, on $\R^{10}$ for example,  if the left- and right- spin structures are
the same, the fermion path integral is manifestly positive, because there are an even number of worldsheet fermions, resulting in a positive semidefinite path integral $(\Pf(\slashed{D}))^{10}$.   So there is no anomaly
in that case.    (More careful arguments lead to the same conclusion with the superghosts included
 and with $\R^{10}$ replaced by a more general spin manifold.)    If the left and right spin structures
are different, it is not so obvious that there  is no anomaly, but this is true.   See  Theorem 4.5 and Corollary 4.6 in \cite{FW}.  The present article does not require
familiarity with those arguments.

For Type IIA, the path integral measure contains an extra factor, namely $(-1)^{\zeta_\ell}$, where $\zeta_\ell$ is the mod 2 index of the left-chiral Dirac operator.
    We would get an equivalent
theory, differing by a  spacetime reflection, if we use $\zeta_r$, the mod 2 index of the right-chiral Dirac operator, instead of $\zeta_\ell$.  To see this, note that a spacetime reflection, say $X^k\to -X^k$ where $X^k$ is
one of the spacetime coordinates, is accompanied (in the RNS formalism of string theory)    by $\psi_k\to -\psi_k$, where $\psi_k$ is the worldsheet superpartner of $X^k$.
The transformation $\psi_k\to -\psi_k$ multiplies every-zero mode of $\psi_k$, regardless of chirality, by $-1$.  So this transformation multiplies the path integral by $(-1)^n$,
where $n$ is the total number of zero-modes of $\psi_k$, regardless of chirality.  (As usual, non-zero modes of $\psi_k$ are paired and do not contribute to the anomaly.)
Since $n=\zeta_\ell+\zeta_r$ mod 2, the spacetime reflection multiplies the path integral by $(-1)^{\zeta_\ell+\zeta_r}$, exchanging a theory with a factor $(-1)^{\zeta_\ell}$
with a theory with a factor $(-1)^{\zeta_r}$.

Concretely, at the 1-loop level,  since the only odd spin structure is of type $++$,
including a factor of $(-1)^{\zeta_\ell}$ will multiply the path integral by $-1$ if the left-moving spin structure is of type $++$, and does nothing in other cases.   Reversing the sign of
the path integral when the left-moving spin structure is of type $++$ has the effect of changing the sign of the  GSO projection for left-moving Ramond modes.   
Reversing the sign of this projection is the usual operation that converts Type IIB superstring theory to Type IIA superstring theory, so we conclude that if we include a factor
$(-1)^{\zeta_\ell}$ in the sum over spin structures, we get the Type IIA superstring theory.

Yet another way to see the relation between Type IIA superstring theory and the factor $(-1)^{\zeta_\ell}$ is the following.    First, compactify the Type II superstring theory on a circle,
parametrized  by one of the spatial coordinates, say $X^k$.
An important result in building up the network of string theory dualities is that $T$-duality on a circle exchanges Type IIA and Type IIB superstring theory \cite{DLP,DHS}.
On the other hand, a relatively well-known
 fact is that $T$-duality on a circle parametrized by $X^k$ has the effect of acting on the corresponding fermion field
 $\psi_k$ by a discrete chiral transformation $\psi_k\to \h\gamma\psi_k$.  As we have
 just learned, this chiral transformation is anomalous and multiplies the path integral by a factor $(-1)^{\zeta_\ell}$.   So $T$-duality on a circle adds or removes a factor
 $(-1)^{\zeta_\ell}$ in the path integral.   By convention, the theory without the factor of $(-1)^{\zeta_\ell}$ is called Type IIB superstring theory, and the one with this
 factor is called Type IIA.   As a quick check on this convention, note that the Type IIB theory and not the Type IIA theory is invariant under reversing the worldsheet
 orientation.   Indeed, the factor $(-1)^{\zeta_\ell}$ is exchanged with $(-1)^{\zeta_r}$ by a reversal of worldsheet orientation, so the theory that is invariant under
 worldsheet orientation reversal is the one without this factor.
 
 \subsection{Branes of Type II Superstring Theory}\label{branestwo}
 
 We are now ready to  discuss branes in Type II superstring theory.
 This means that we want to define the theory on a Riemann surface with boundary with some sort of boundary condition.
   Let us start with Type IIB.    In the case of a D9-brane, all worldsheet fields $X^k$ that represent target space coordinates satisfy the same boundary
   condition (namely Neumann), and therefore superconformal symmetry requires that the superpartners $\psi_k$ also satisfy a common boundary condition,
   which takes the form  
\be\label{indico}\bigl.\vec\gamma\cdot \vec n\psi\bigr|=\bigl.\varepsilon\psi\bigr|,\ee
(where $\vec n$ is the unit normal to the boundary and $\bigl.\psi\bigr|$ is the restriction of $\psi$ to the boundary) with  $\varepsilon=\pm 1$.     The overall choice of $\varepsilon$ is inessential in that it can be reversed by a chiral rotation $\psi\to\h\gamma\psi$.

To describe in Type IIB a Dp-brane, we flip the boundary condition of $9-p$ boson fields $X^k$ (the ones that parametrize directions normal to the brane) from Neumann
to Dirichlet.   Superconformal symmetry then requires that we also flip the boundary condition for the corresponding 
$9-p$ worldsheet fermions.      If $9-p$ is even, this introduces no anomaly and we get the
usual supersymmetric Dp-branes of Type IIB superstring theory with odd $p$.    
If $9-p$ is odd, flipping the boundary conditions for that many fermions introduces an anomaly.    But we can cancel the anomaly
by adding another real fermion $\lambda$ that only propagates on the boundary.      This gives the nonsupersymmetric Type IIB Dp-branes of \cite{Sen1,Sen2,Sen3} with even 
$p$.    We will say more about them presently.

Now let us consider Type IIA.     Here we are in the opposite situation.      If we try to make a D9-brane, then since the fermions all satisfy the same boundary condition,
they introduce no anomaly.     But for Type IIA, there is a bulk factor $(-1)^{\zeta_\ell}$, and this has an anomaly on a Riemann surface with boundary.     To cancel the anomaly,
one way is to flip the sign of the boundary condition for an odd number of bulk fermions.    To maintain superconformal symmetry, we should also flip the boundary condition
for the same number of bosons, by considering a Dp-brane with $9-p$ odd and thus $p$ even. In this way, we arrive at  
 the usual supersymmetric
Type IIA Dp-branes with even $p$.     Alternatively, to get a Dp-brane with odd $p$, we have to flip  the boundary condition for an even number of bulk fermions.     This will not
affect the anomaly, but we can cancel the anomaly by adding a real fermion $\lambda$ that only propagates on the boundary.   This gives  nonsupersymmetric Dp-branes of Type IIA
with $p$ odd that were constructed and analyzed in \cite{Sen1,Sen2,Sen3}.

Finally let us discuss the physics of these nonsupersymmetric branes.     We can quickly reproduce many of their slighly unusual properties.   First is the brane tension.
In general, the tension of a brane is computed from the path integral on a disc (or equivalently from the disc contribution to the dilaton one-point function) with boundary condition
set by that brane.  
 The path integral on a disc gets a factor of $\sqrt 2$ from the boundary fermion, as explained in section \ref{threety}.    So the nonsupersymmetric Dp-brane tension has an extra
factor $\sqrt 2$ compared to a supersymmetric brane of the same $p$ (which would exist in the opposite string theory).   This is a result in \cite{Sen1,Sen2,Sen3}.

Next we can consider vertex operators.    What sort of vertex operator $V$ can we insert on a boundary of the string worldsheet associated to a nonsupersymmetric brane?
 Usually, $V$ is a superconformal primary of appropriate dimension constructed from the available matter fields.   In addition, it is constrained to be GSO even. 
In the case of a nonsupersymmetric brane, however, we also have the dimension 0 fermion  field $\lambda$  that is defined only on the boundary.   So if $V$ is GSO-odd, then
$\lambda V$ is GSO-even.    This means that if we do not explicitly take $\lambda$ into account, we would say that there is no GSO projection: for every vertex operator $V$
(for simplicity with definite behavior under GSO) either $V$ or $\lambda V$ is GSO-even.   This is again part of the story in \cite{Sen1,Sen2,Sen3}.

 \begin{figure}
 \begin{center}
   \includegraphics[width=2.5in]{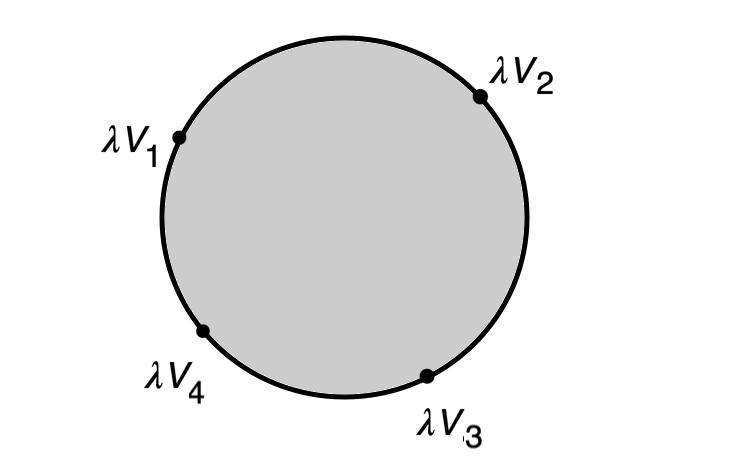}
 \end{center}
\caption{\footnotesize  On a worldsheet boundary labeled by a ``wrong parity'' Dp-brane,  the number of vertex operators insertions of 
the type $\lambda V$ (in other words, insertions of vertex operators that are GSO-odd if one does not take the boundary
fermion $\lambda$ explicitly into account) 
must be even (odd) in the case of a boundary of NS (Ramond) type, in order to avoid
vanishing of the $\lambda$ path integral.  The NS case is depicted here. \label{four}}
\end{figure} 

There is also a selection rule for the insertion of GSO-odd vertex operators, that is GSO-even vertex operators of the form $\lambda V$ where $V$ is GSO-odd. 
The statement depends on whether $\lambda$ has a zero mode.    On a boundary component on which the spin structure is of NS type, $\lambda$ has no zero-mode, and hence
the number of GSO-odd vertex operator insertions must be even (fig. \ref{four}).
 But on a circle on which the spin structure is of Ramond type, there is a $\lambda$ zero-mode, and the number of GSO-odd insertions must be odd.  Again this is
part of the story in \cite{Sen1,Sen2,Sen3}. 

Pick a particular nonsupersymmetric D-brane  $\B$.   Its boundary action contains a Majorana fermion $\lambda$ with the action (\ref{zelbo}).  Define an operator $(-1)^\lambda$ that assigns the value $+1$ to  vertex operators of the $\B$-$\B$ system of type $V$ and $-1$ to those
of type $\lambda V$. Thus $(-1)^\lambda$ corresponds to the operation $\lambda\to-\lambda$, which classically is a symmetry of the action.   The selection rule just described implies that the path integral on a worldsheet $\Sigma$ violates conservation of $(-1)^\lambda$ if and only if $\Sigma$
has an odd number of Ramond type boundary components labeled by $\B$.     We can describe this by saying that Ramond boundaries produce an anomaly in the classical
symmetry $(-1)^\lambda$.
We will discuss the consequences in section \ref{antibranes}. 

Because there is no GSO projection on a nonsupersymmetric Type II Dp-brane, the nonsupersymmetric Dp-branes of Type IIA and Type IIB are all tachyonic -- there is always
a tachyon vertex operator in the Dp-Dp sector.    If  $k\cdot \psi \exp(\i k\cdot X)$ denotes (the top component of) the usual chiral GSO-odd tachyon vertex operator,
then the GSO-even tachyon vertex operator of a nonsupersymmetric Dp-brane is $\lambda k\cdot \psi \exp(\i k\cdot X)$.

\subsection{Branes and Antibranes}\label{antibranes}

A D-brane is characterized in part by the boundary conditions satisfied by worldsheet bosons and fermions.  Consider a D-brane $\B$ supported on a submanifold $W$ in a spacetime $M$.
On a worldsheet boundary labeled by $\B$, worldsheet bosons $X$ that parametrize $W$ satisfy Neumann boundary conditions, and worldsheet bosons $Y$ that parametrize
the normal directions satisfy 
Dirichlet boundary conditions.    Related to this, the fermionic partners $\psi_X$ of $X$ satisfy a boundary condition with one sign, and the partners $\psi_Y$ of $Y$ satisfy
a ``flipped'' boundary condition with the opposite sign.

This is the way we have described supersymmetric 
D-branes so far, but it misses a key aspect:  the distinction between D-branes and ${\overline {\mathrm D}}$-branes.    A supersymmetric D-brane is a source of a Ramond-Ramond
field.   In the case that $W$ has codimension $m$, the Ramond-Ramond field can be described as an $(m-1)$-form $G$ that is magnetically coupled to $W$, satisfying
\be\label{elbow}\d G=\delta_W, \ee
where $\delta_W$ is an $m$-form delta function supported on $W$.   If $W$ is defined locally by conditions $Y^1=\cdots = Y^m=0$, with normal coordinates $\vec Y$,
then
\be\label{lbow}\delta_W=\delta(\vec Y)\d Y^1\d Y^2\cdots \d Y^m. \ee
To define the sign of the $m$-form $\d Y^1\d Y^2\cdots \d Y^m$ that appears in this formula requires an orientation of the normal bundle $N$ 
to $W$ in $M$.   (If $M$ itself is oriented, which is
the case in Type IIB superstring theory, then an orientation of the normal bundle is equivalent to an orientation of $W$.)   Reversing the orientation of $N$ 
will reverse the sign of $\delta_W$ and therefore of the Ramond-Ramond field $G$ that is sourced by the brane.   This
 is equivalent to replacing a D-brane wrapped on $W$ with a ${\overline {\mathrm D}}$-brane.

The distinction between supersymmetric D-branes and ${\overline{\mathrm  D}}$-branes, or equivalently the orientation of  the normal bundle to 
$W$, is not encoded in the bosonic and fermionic boundary conditions
that characterize the brane.   To understand it, we must look more closely at the $\Z_2$-valued anomaly of the worldsheet path integral.

Let us first discuss why $M$ itself must be oriented in Type IIB superstring theory.  We consider a worldsheet $\Sigma$ without boundary and we choose an odd spin structure for negative chirality fermions and an even one for those of positive chirality.  In superstring perturbation
theory, one integrates over maps $\Phi:\Sigma\to M$; for simplicity, consider the case\footnote{\label{consider} By considering general maps $\Phi$, not  necessarily constant,
one  learns that $M$ must be spin, not just oriented.} that $\Phi$ maps $\Sigma$ to a point $p\in M$.   Each of the ten worldsheet fermions $\psi^I$, $I=
1,\cdots,10$ has an odd number of zero-modes -- generically precisely 1.   Let us call the zero-modes $\eta^1,\cdots,\eta^{10}$.   To define the path integral with this choice of spin
structures, we need a measure for the zero-modes.   This measure will be
 a multiple of $\d \eta^1\cdots \d \eta^{10}$, but there is no natural choice of the sign of this measure.  To choose the sign
of the measure requires an orientation of the tangent space to $M$ at $p$, telling us how to order the $\eta$'s up to an even permutation.
  If $M$ is unorientable, the theory is anomalous: the sign of the fermion measure will be reversed when
$p$ moves continuously around an orientation-reversing loop in $M$.   If $M$ is orientable, there is no anomaly and the measure can be defined consistently, but to actually define
the measure, we need to pick an orientation on $M$.

Once we have done so, we then have to decide what to do in the opposite case that
 the spin structure is even for negative chirality fermions and odd for those of positive chirality.   One option is to make
the same choice for both chiralities.  Doing this, we get a theory that is invariant under orientation-reversal of the worldsheet, but whose definition depends on an orientation
of $M$. This is called Type IIB superstring theory.  If we make opposite choices for the negative and positive chirality fermions, we get a theory which does not require
an orientation of spacetime; the spacetime orientation can be reversed if one also reverses the worldsheet orientation.   This is called  Type IIA superstring theory.  This explanation
of the relation between Type IIA and Type IIB also implies that multiplying by, say, $(-1)^{\zeta_r}$ will exchange these two theories, as was explained already in section \ref{super}.

This example gives a simple illustration of the difference between an anomaly being trivial and the anomaly being trivialized.   If $M$ is unorientable, then Type IIB superstring theory
on $M$ is anomalous -- there is no way to define consistently the sign of the worldsheet path integral with target $M$.   If $M$ is orientable, then the theory is anomaly-free and it is
possible to define the sign consistently, but for certain spin structures,  there is no preferred choice of sign -- to define the sign requires choosing an orientation of $M$. 
When we choose an orientation, we are trivializing the anomaly.
  Reversing the orientation of $M$
will reverse the sign of the path integral measure if the negative or positive chirality fermions, but not both, have an odd spin structure.  That sign reversal  can be accomplished by
 multiplying the
path integral by $(-1)^{\zeta_\ell+\zeta_r}$.  The fact that in Type IIB superstring theory,
 multiplying the path integral measure by that factor is equivalent to reversing the spacetime orientation was explained in a different but related way in section \ref{super}.
 
 The need to orient the normal bundle to a D-brane comes from a more subtle variant of these considerations.   In either Type IIA or Type IIB superstring theory, 
 consider a nonsupersymmetric D-brane $\B$ supported on $W\subset M$.
 Consider a worldsheet $\Sigma$ with a boundary component $S$ that is labeled by $\B$, so that $\Phi(S)\subset W$.  Along $S$, the boundary condition of the fermionic
 partners of normal coordinates $Y^1,\cdots, Y^m$  is flipped in sign.   
 From section \ref{threety}, we know that as far as the problem of defining the sign of the fermion path integral is concerned, this flip of the boundary condition has the same
 effect as adding Majorana fermions $\chi^1,\cdots, \chi^m$ that propagate only along $S$ and are in $1-1$ correspondence with the $Y^i$.  The $\chi^i$ do not actually
 exist in Type II superstring theory, but flipping the boundary condition for the superpartners of the $Y^i$  has the same effect on the 
 problem of defining the fermion measure as adding the $\chi^i$, and the following considerations are more obvious if we think in terms of the $\chi$'s.   
 Also, we will simplify the problem by considering only the case that the map $\Phi:\Sigma\to M$ is constant\footnote{Analogously to the remark in footnote \ref{consider},
 analyzing the general case leads to the condition that
 the ``$\UU(1)$ bundle'' along the support of a supersymmetric Type II D-brane is really a ${\mathrm{Spin}}_c$ structure \cite{FW}. This is an aspect of the 
 relation between D-branes and K-theory.}  when restricted to $M$.   In this case, if and only if the spin structure of $S$ is of Ramond type, each of the $\chi^i$ has a single zero-mode $\eta^i$, and to define the path integral measure for the
 $\chi^i$ requires choosing the sign of the measure $\d\eta^1\d\eta^2\cdots\d\eta^m$.  
 
  Up to this point, in our study of nonsupersymmetric D-branes,
 we considered only an anomaly in $(-1)^\fF$, and we observed that a measure $\d\eta^1\d\eta^2\cdots \d\eta^m$ is invariant under $(-1)^\fF$ if and only if $m$ is even.
   But there is more to say; we can also ask what happens to the measure if we make a linear transformation of the $Y^i$.   
  Since the $\chi^i$ and therefore the $\eta^i$ are in natural correspondence
 with the $Y^i$, an orientation-reversing linear transformation of the $Y^i$ will reverse the sign of the measure $\d \eta^1\d\eta^2\cdots\d\eta^m$.  
 Therefore, in general the sign of the fermion path integral in the presence of a D-brane supported on $W$ cannot be defined consistently if the normal bundle $N$ to $W$
 is unorientable.   If $N$ is orientable, then the sign of the path integral can be defined consistently, but to define the sign, we have to pick an orientation of $N$.
 
 To see the implications, let $V_G$
 be the vertex operator of the Ramond-Ramond field $G$.
The $G$-field sourced by $\B$ is computed by
 the one-point function of $V_G$ on a disc with boundary
 condition set by $\B$.   The boundary of the disc carries a Ramond spin structure.   So this is precisely the situation in which the sign of the path integral depends on the orientation 
 of the normal bundle to $W$ in $M$. Reversing the orientation changes the sign of the one-point function.   Thus the distinction between supersymmetric D-branes and 
 ${\overline{\mathrm D}}$-branes enters in giving a precise definition of the sign of the fermion path integral.  
 
 A way to summarize all this is to say that, given the definition of the worldsheet path integral for a D-brane $\B$, the worldsheet path integral for the corresponding antibrane $\bar\B$
 can be defined by just including a factor of $-1$ for every Ramond boundary labeled by $\B$.   Here the relation between $\B$
 and $\bar\B$ is actually perfectly symmetrical.  It is equally  true that the worldsheet path integral for $\B$ can be obtained  from that for $\bar\B$ by including a factor $-1$
 for every Ramond boundary labeled by $\bar\B$.
 
 We have explained in detail  these matters, which are certainly not essentially novel,
   in the hope of making it obvious that there is {\it not} a similar distinction between branes and antibranes in the case of a nonsupersymmetric
 Type II D-brane.   Consider a nonsupersymmetric D-brane $\B$ with support $W$, and let $S$ be a worldsheet boundary  labeled by $\B$.   Along $S$, there propagates a Majorana fermion
 $\lambda$.  In the previous setup that led to zero-modes $\eta^1,\cdots,\eta^m$ in $1-1$ correspondence with the $Y^i$, there is now one more zero-mode, the zero-mode 
 of $\lambda$, which we will call $\eta^0$.   
 So instead of trying to define a measure $\d \eta^1\cdots \d \eta^m$, we now want to define a measure $\d\eta^0\d\eta^1\cdots \d\eta^m$.    There is no problem in
 defining the measure, even for  unorientable $N$, if we say that $\lambda$ and therefore $\eta^0$ is odd under an orientation-reversing linear transformation of the $Y$'s.
 More intrinsically, we should slightly refine our definition of a nonsupersymmetric D-brane by saying that $\lambda$ is a Majorana fermion valued in a real line bundle
 which is the orientation bundle of $N$.   Then there is a completely canonical measure  $\d\eta^0\d\eta^1\cdots \d\eta^m$.
 
 To see what is going on in a possibly more down-to-earth way, let us see what happens if we imitate the operation that distinguishes supersymmetric branes and antibranes.  Suppose
 that we modify  the worldsheet path integral in the presence of a nonsupersymmetric D-brane $\B$ by
 including a factor $-1$ for every Ramond boundary labeled by $\B$.   The selection rule explained in section \ref{branestwo} shows that this has the same effect as acting
 with $(-1)^\lambda$ on the vertex operators.  Putting this differently,  including the factor $-1$ for every Ramond boundary labeled by $\B$ can be compensated by acting with $(-1)^\lambda$ on all vertex
 operators.   The situation is analogous to what happens in QCD, where in the presence of a quark with zero bare mass, there is an anomalous chiral symmetry by virtue of which
 theories with different values of the QCD theta-angle are actually equivalent.   Here,  branes $\B,\B'$ that differ only by a factor $-1$ for every Ramond boundary labeled by $\B$ are equivalent
 because of the anomalous symmetry $(-1)^\lambda$.
 
 The same occurs in Type I superstring theory.   As we will see, all nonsupersymmetric Type I
 D-branes are constructed with at least one Majorana fermion propagating on the worldsheet boundary.
 This leads to the existence of an anomalous symmetry, similar to $(-1)^\lambda$ and ensuring that there is no distinction between nonsupersymmetric D-branes and $\overline{\mathrm D}$-branes.

\section{Branes in Type I Superstring Theory}\label{TypeI}

We now turn to Type I superstring theory.

\subsection{Time-Reversal Symmetry and the Orientifold Projection}\label{time}

 \begin{figure}
 \begin{center}
   \includegraphics[width=3.5in]{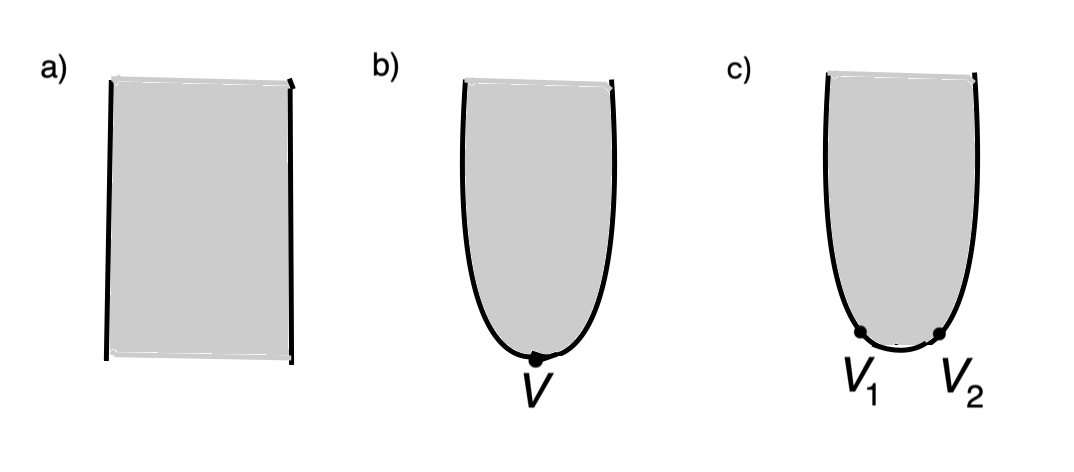}
 \end{center}
\caption{\footnotesize  (a) The orientifolding symmetry is a reflection $\rR$ that  exchanges the two ends of an open string.   (b)   Introducing a vertex operator
$V$ that creates the open string state in question, $\rR$  leaves fixed the point $p$ at which $V$ is inserted and reverses the boundary orientation
near  $p$.     (c) The reflection $\rR$ reverses the order with which vertex operators $V_1$, $V_2$ are inserted on the boundary.     \label{five}}
\end{figure} 

Type I superstring theory can be viewed as an orientifold of a Type IIB superstring theory with D9-branes.   In general, orientifolding is achieved by dividing by a symmetry 
 that reverses the orientation of a string worldsheet, exchanging the two ends (fig. \ref{five}(a)).  
 This is often called $\Omega$, but here we will call it $\rR$ to emphasize that it is a spatial reflection symmetry.

The $\sf{CPT}$ theorem says that if and only if a theory has a spatial reflection $\rR$, it also has a time-reversal symmetry $\tT$.   (For some general comments
on the $\sf{CPT}$ theorem and the associated terminology, see Appendix \ref{cpt}.)   This can be understood as follows. In Euclidean signature, in a quantum
field theory on $\R^n$, with coordinates $(x_1,x_2,\cdots,x_n)$, consider a reflection symmetry $\rR$ that acts by $(x_1,x_2,\cdots,x_n)\to (-x_1,x_2,\cdots,x_n)$.
There are two essentially different ways to Wick rotate this theory to Lorentz signature.  If we Wick rotate one of the coordinates $x_2,\cdots , x_n$ that are not reflected by $\rR$,
then $\rR$ remains as a unitary reflection symmetry in Lorentz signature.    On the other hand, if we Wick rotate the coordinate $x_1$ that is reflected by $\rR$,
then $\rR$ continues in Lorentz signature to an antiunitary time-reversal symmetry $\tT$.   Thus a relativistic 
quantum field theory has a reflection symmetry if and only if it has a time-reversal 
symmetry.   

Thinking about the vertex operator $V$ that creates a given open-string state (fig. \ref{five}(b)), we see that $\rR$ can be naturally defined to leave fixed the point $p$ at
which $V$ is inserted, while reversing the orientation of the boundary (and bulk) of the string worldsheet near $p$.   Given two or more vertex operators inserted on the same
boundary, $\rR$ reverses the order in which they are multiplied, that is, in which they are inserted on the boundary  (fig. \ref{five}(c)).

In open-string theory,  the bulk worldsheet theory is often tensored with a $0+1$-dimensional topological field theory
that lives only on the boundary of the open string.   A field theory in $0+1$ dimensions is just a quantum mechanics system, and to say that it is topological just means
that its Hamiltonian vanishes.  The familiar boundary quantum mechanics system in open-string theory is associated to a Chan-Paton gauge group: 
one attaches to a boundary of an open-string a finite-dimensional Hilbert space $\H_b$, with zero Hamiltonian, and then the 
operator algebra of the theory is extended by including  operators that act on $\H_b$.   Such operators, in particular, can be used in constructing vertex operators.  
For our purposes in this article,  there is another important example of a boundary quantum mechanics system: one can have Majorana fermions propagating
on the boundary.

To understand vertex operators that can be inserted on a worldsheet boundary labeled by a given brane in Type I superstring theory, we need to understand the action of $\rR$ or $\tT$ on the 
boundary quantum mechanics.   When one thinks of a vertex operator as an insertion in a correlation function on a Euclidean worldsheet, $\rR$ is often
the more natural symmetry.   But when one thinks of a vertex operator as an operator that acts on a Hilbert space, $\tT$ is often more natural.   
  So  it is useful  to understand
how to relate the two symmetries.

    We will denote as $\A$ the algebra of operators of a boundary
  quantum mechanics system; $\a,\b,\cdots$ will denote elements of $\A$.   A reflection $\rR$ acts linearly on boundary operators in general, so in particular it acts
  linearly on operators of the boundary quantum mechanics, 
say by $\a\to \rR(\a)$.
But as we see in fig. \ref{five}(c), $\rR$ 
reverses the order in which operators are inserted on the boundary, so 
\be\label{topo}\rR(\a\b)=\rR(\b)\rR(\a). \ee
Mathematically, this means that $\rR$ is an isomorphism from the algebra $\A$ of local operators of the boundary quantum mechanics to its 
opposite algebra $\A^\op$. The definition of $\A^\op$ is the following:
elements $\a^\op$ of $\A^\op$ are in $1-1$ correspondence with elements $\a$ of $\A$, but they are multiplied in the opposite order: $(\a\b)^\op =\b^\op\a^\op$.

In contrast to $\rR$, $\tT$ is an antilinear map from the boundary algebra $\A$ to itself.  However, it does not reverse the order in which operators are multiplied.
Rather,
\be\label{opo} \tT(\a\b)=\tT(\a)\tT(\b).   \ee
To see this, we simply observe that in a Hilbert space realization, we can view $\A$ as the algebra  of boundary operators acting on, say, the left end of the string, while
$\tT$ is an antilinear operator on the Hilbert space that maps the left (or right) end of the string to itself; in this formulation, $\tT(\a)=\tT\a\tT^{-1}$, from which  (\ref{opo}) follows.   We cannot make this argument for $\rR$,
because although $\rR$ can indeed be viewed as a Hilbert space operator, as such it exchanges the left and right ends of the string and hence conjugates the algebra $\A$
of boundary operators on the left end of the string to a similar algebra on the right end.   To interpret $\rR$ as a map from an algebra $\A$ of boundary operators to itself,
we have to use the Euclidean picture of figs. \ref{five}(b,c), and in that picture it is not true that $\rR(\a)$ can be interpreted as $\rR \a \rR^{-1}$.   

The $\sf{CPT}$ theorem is therefore telling us that there is a natural correspondence between linear isomorphisms $\rR:\A\to\A^\op$ and antilinear isomorphisms
$\tT:\A\to \A$.   This statement may seem puzzling at first, but it actually follows from the existence of the operation of hermitian conjugation of boundary operators:
$\a\to\a^\dagger$.  As hermitian conjugation is antilinear and $(\a\b)^\dagger=\b^\dagger\a^\dagger$,  hermitian conjugation is an
antilinear isomorphism from $\A$ to $\A^\op$.   Having an antilinear isomorphism from $\A$ to $\A^\op$ does indeed give a natural way to convert linear isomorphisms from $\A$
to $\A^\op$ into antilinear isomorphisms from $\A$ to $\A$.   Given $\rR$ or $\tT$, we define the other one by 
\be\label{belko} \rR(\a)=(\tT(\a))^\dagger, ~~~~~\tT(\a)=(\rR(\a))^\dagger. \ee
This is the relation between the automorphisms of the boundary algebra associated with $\rR$ and $\tT$.  

The conclusion (\ref{belko}) actually is valid more generally for boundary operators
whose definition involves the bulk matter fields, not only for operators of a boundary quantum mechanics system.  Generic boundary operators 
generate not an ordinary algebra but a more subtle operator product algebra, but eqn. (\ref{belko}) still holds.

We will work out the implications of eqn. (\ref{belko}) in detail, first for the classic case of real or quaternionic Chan-Paton factors, and then for the case of boundary fermions that is 
important in understanding
nonsupersymmetric D-branes.  

To get real Chan-Paton factors, we introduce a boundary Hilbert space $\H_b$ of dimension $n$ on which $\tT$ acts as complex conjugation; we denote this by $\tT=\star$.
$\A$ is then the algebra of $n\times n$ matrices acting on $\H_b$; we denote the complex conjugate and transpose of a matrix $\a$ as $\bar\a$ and $\a^\tr$, respectively.
We have  $\tT(\a)=\tT \a \tT^{-1}=\bar\a$, from which it follows that
\be\label{rff}\rR(\a)=(\tT(\a))^\dagger =\a^\tr. \ee
   Now let us see why this leads to an orthogonal gauge group.   Parametrizing the worldsheet boundary  by $\tau$, the open-string
vertex operator\footnote{The vertex operators that we will encounter are in supermultiplets, and we describe them by giving the top component of the supermultiplet.} 
describing a gauge field of momentum $k$ and polarization $\epsilon$  is 
\be\label{newg} V_\gauge=\epsilon_I \left(\partial_\tau X^I +\i k_J\psi^I\psi^J \right)\exp(\i k\cdot X). \ee
This is odd under $\rR$ ($\partial_\tau X$ is odd because $\rR$ reverses the sign of $\tau$, and $\psi^I\psi^J$ is odd because $\rR$ reverses the order in which these
two fermion operators are inserted on the boundary).   An $\rR$-invariant vertex operator  $V$ describing a gauge field  will have the form $V=\a V_\gauge$ where $\a\in \A$ is also odd,
to compensate for the oddness of $V_\gauge$.  Since $\rR(\a)=\a^\tr$,
$\a$ is odd if and only if it is antisymmetric.  Antisymmetric matrices generate the Lie algebra  $\mathfrak{so}(n)$, so this is the Lie algebra of the gauge group in this situation.

The minimal construction that gives symplectic Chan-Paton factors is to take $\H_b$ to be a vector space of dimension $2$ on which $\tT$ acts by
\be\label{nff}\tT=\star M, \hskip1cm M=\begin{pmatrix} 0 & -1 \cr 1 & 0 \end{pmatrix}.\ee
The basic difference between this case and the previous one is that with this choice, $\tT^2=-1$.   If $\tT^2=1$, then up to a unitary transformation, one can assume $\tT=\star$, as 
before,
and if $\tT^2=-1$, then up to a unitary transformation, $\tT$ has a block diagonal form with each block of the form (\ref{nff}).  Eqn. (\ref{nff}) leads to
\be\label{piff}\tT(\a)=M\bar \a M^{-1}, 
~~\rR(\a)= M\a^\tr M^{-1}.\ee
Explicitly, if $\a=a_0+\vec \sigma\cdot \vec a$ is the expansion of $\a$ in the basis $1,\vec\sigma$ of $2\times 2$ matrices ($\vec \sigma$ are the Pauli matrices), then $\rR(\a)=a_0-\vec\sigma\cdot \vec a$.  For a vertex operator $\a V_\gauge$
to be even under $\rR$,  $\a$ must be odd and therefore must have the form $\a=\vec\sigma\cdot\vec a$.  Such matrices make up the Lie algebra $\mathfrak{su}(2)$, and so that is the Lie algebra of  the gauge group in this situation.  A slight extension of this analysis shows that  if we take $\H_b$ to be of dimension $2n$ with $\tT$ the direct sum of $n$ blocks each of the form (\ref{nff}),  then the gauge
algebra  is $\mathfrak{sp}(n)$  (in conventions such that $\mathfrak{sp}(1)=\mathfrak{su}(2)$).  Thus $\tT^2=1$ and $\tT^2=-1$ lead to Chan-Paton constructions of orthogonal and
symplectic gauge groups, respectively.

In this article, an important example is that the boundary theory contains $k$ Majorana fermions $\lambda_1,\cdots, \lambda_k$, which may be even or odd under $\tT$:
$\tT(\lambda_i)=\veps_i \lambda_i$ with $\veps_i=\pm 1$.    Then the action of $\tT$ on a product of $\lambda$'s is
\be\label{omigo}\tT(\lambda_{i_i}\cdots \lambda_{i_r})=\lambda_{i_i}\cdots \lambda_{i_r}\cdot \prod_{j=1}^r\veps_{i_j} \ee
and $\rR$ acts in the same way, except that it reverses the order in which the $\lambda$'s are multiplied:
\be\label{omigot}\rR(\lambda_{i_i}\cdots \lambda_{i_r})=\lambda_{i_r}\cdots \lambda_{i_1}\cdot \prod_{j=1}^r\veps_{i_j} .\ee

Having understood the relationship between $\tT$ and $\rR$, we can study boundary theories in a Hamiltonian framework, where $\tT$ is more natural,
and carry over the results to vertex operators in Euclidean signature, where $\rR$ is more natural.

\subsection{Time-Reversal and Majorana Fermions}\label{majotime}

With time-reversal symmetry in mind, we reconsider the theory of a single Majorana fermion $\lambda $ that propagates only on the worldsheet boundary, with action
\be\label{wac}I=\int\d \tau \frac{1}{2}\i \lambda \frac{\d\lambda}{\d \tau}. \ee
This theory actually has both 
a unitary symmetry $(-1)^\fF$ that acts by $\lambda\to -\lambda$, and an anti-unitary  time-reversal symmetry $\tT$ that acts by $\tT\lambda(\tau)\tT^{-1}=\lambda(-\tau)$.
Since the Hamiltonian vanishes, $\lambda$ is actually independent of $\tau$ and we can abbreviate that formula.
\be\label{simac} \tT \lambda\tT^{-1} =\lambda.\ee
Classically the operators $(-1)^\fF$ and $\tT$   obey 
\be\label{iwac}\bigl((-1)^\fF\bigr)^2=\tT^2=1,~~~~ (-1)^\fF\tT = \tT (-1)^\fF. \ee
If a theory that classically has $(-1)^\fF$ and $\tT$ symmetries
can be quantized in such a way that in an irreducible representation of the operator algebra, unitary and anti-unitary operators $(-1)^\fF$ and $\tT$
can be defined that transform $\lambda$ in the expected way and satisfy (\ref{iwac}),  then we say that the theory is anomaly-free; if not, then we call it anomalous.

A preliminary comment is that we could have taken $\lambda$ to transform under $\tT$ by
$\tT\lambda \tT^{-1} = -\lambda$, with an extra minus sign. We say that $\lambda$ is $\tT$-even or $\tT$-odd if the sign on the right hand side of this formula is $+$ or $-$.
  However, as long as there is only one Majorana field $\lambda$, and more generally as long as all such fields transform with
the same sign under $\tT$, reversing the sign with which $\tT$ acts on $\lambda$ does not add anything essentially new, as it amounts to replacing $\tT$ with $\tT'=\tT (-1)^\fF$.  The operators $\tT'$ and $(-1)^\fF$ generate
the same group as $\tT$ and $(-1)^\fF$.   When we construct Type I superstring theory, $\tT$ and $(-1)^\fF$ generate constraints (the orientifold and GSO projections,
respectively), and it does not matter what generators we pick of the group of constraints.

  The sign in the action of $\tT$ does become meaningful, however, in a theory that has some $\tT$-even Majorana fermions and some that are $\tT$-odd.  
The basic case is a pair of Majorana fields $\lambda$ and $\t\lambda$ that transform with opposite signs under $\tT$, say $\tT \lambda  \tT^{-1}=\lambda$,
$\tT \t\lambda\tT^{-1}=-\t\lambda$.    A theory of such a pair is completely anomaly-free.   To see this, note that we can represent the fields $\lambda$, $\t\lambda$
by $2\times 2$ real Pauli matrices, say $\lambda=\sigma_1$, $\t\lambda=\sigma_3$. This satisfies the Clifford algebra $\lambda^2=\t\lambda^2=1$, $\{\lambda,\t\lambda\}=0$.
  Then defining $(-1)^\fF=-\i \sigma_1\sigma_3=\sigma_2$, and
$\tT=\star\sigma_1$, we find that $\lambda$ and $\t\lambda$ transform as desired and all conditions (\ref{iwac}) are satisfied.

Thus in investigating anomalies, we can ``cancel'' a pair of Majorana modes that transform with opposite signs under $\tT$.   The basic case to consider therefore is a collection of $N$
Majorana fermions that all transform with the same sign under $\tT$, say with a plus sign:   $\tT \lambda_i\tT^{-1} = +\lambda_i$, $i=1,\cdots,N$.

For what values of $N$ is this theory anomaly-free?\footnote{The following material has been explained in many places, for example in \cite{SW}.}   Even if we ignore the time-reversal symmetry, we know that the theory is anomalous if $N$ is odd.   It turns out that taking into
account $\tT$ as well as $(-1)^\fF$, the theory of $N$ Majorana fields in $0+1$ dimensions, all transforming the same way under $\tT$, is anomaly-free if and only if $N$ is a multiple of 8
\cite{FK,ABS}.   

To understand this, it is useful to first show that for any even $N=2k$, one can find an irreducible representation of $N$ gamma matrices in a Hilbert space of dimension $2^{k}$ such
that $k$ of the gamma matrices are real and $k$ are imaginary.  For $N=2$, we introduce a single qubit (a system with a two-dimensional Hilbert space) and take
\be\label{first}\gamma_1=\sigma_1,~~~\gamma_2=\sigma_2.  \ee
For $N=4$, we add a second qubit, with $\gamma_1$ and $\gamma_2$ acting on the first qubit as before and $\gamma_3,\gamma_4$ acting on both qubits:
\be\label{second}\gamma_1=\sigma_1\otimes 1, ~~\gamma_2=\sigma_2\otimes 1,~~~~\gamma_3=\sigma_3\otimes \sigma_1, ~~~\gamma_4=\sigma_3\otimes \sigma_2.\ee
Each time we increase $N$ by 2, we add another qubit on which the ``old'' gamma matrices act trivially, and we add two ``new'' gamma matrices that act as $\sigma_3$
on the ``old'' qubits and as $\sigma_1$ or $\sigma_2$ on the new one:
\be\label{third}\gamma_{N-1}=\sigma_3\otimes\cdots  \otimes \sigma_3\otimes \sigma_1,~~~\gamma_{N}=\sigma_3\otimes \cdots \otimes \sigma_3\otimes \sigma_2.\ee
Thus $\gamma_p$ is real or imaginary depending on whether $p$ is odd or even.   

If and only if $N$ is divisible by 8, $\Gamma=\gamma_2\gamma_4\gamma_6\cdots \gamma_N$ is  real and satisfies $\Gamma^2=1$.   When this is the case,
we can define a new set of gamma matrices, all of them real, by setting $\gamma'_{2p+1}=\gamma_{2p+1}$, $\gamma'_{2p}=\i\gamma_{2p}\Gamma$.   This
being so, we can quantize the theory with $\lambda_k=\gamma'_k$, and set $(-1)^\fF=\gamma'_1\gamma'_2\cdots \gamma'_N$, $\tT=\star$. All conditions are satisfied, so  
 there is no anomaly if $N$ is a multiple of 8.  

Let us now investigate the other cases of even $N$.   For any even $N$, with $\lambda_k=\gamma_k$, we can define an operator $(-1)^\fF$ that anticommutes with all $\lambda_k$ 
and whose square is 1 by setting
\be\label{zelb}(-1)^\fF=\i^{N(N-1)/2} \gamma_1\gamma_2\cdots\gamma_N.\ee
Similarly, we can  define a time-reversal symmetry that commutes appropriately with the
$\lambda_k$ by setting 
\be\label{timerev}\tT=\begin{cases} \star \gamma_2\gamma_4\cdots \gamma_{N} ~~~~~~~&N\cong 0 ~{\mathrm{mod}} ~4 \cr
                                                         \star \gamma_1\gamma_3\cdots \gamma_{N-1}~~~~~~~&N\cong 2~{\mathrm{mod}}~4.\end{cases}\ee
But we only get $\tT^2=1$, $\tT(-1)^\fF=(-1)^\fF\tT$ if $N$ is divisible by 8.   For $N\cong 2,6~{\mathrm{mod}}~8$, we find $\tT(-1)^\fF=-(-1)^\fF\tT$,
and for $N\cong 4,6$ mod 8, we get $\tT^2=-1$.       The four cases of even $N$ are distinguished by the signs in these formulas, as summarized in Table \ref{tabull}.

  \begin{table}
\begin{center}
\begin{tabular}{ |c|c|c| c|c |} 
 \hline
  $N~{\mathrm {mod}}~8 $&0 & 2&4&6  \\ \hline
 $\tT^2$ & 1  & $1$ &$-1$&$-1$ \\ \hline
 $\alpha$ & 1 & $-1$ & 1 &$-1$ \\ 
 \hline
\end{tabular}
 \caption{\label{tabull} This table summarizes the anomalies for even $N$ mod 8. $\alpha$ is the sign that appears in $\tT(-1)^\fF=\alpha(-1)^\fF\tT$.  The four cases of even
 $N$ mod 8 are distinguished by $\alpha$ and the sign of $\tT^2$.}   
\end{center}
\end{table}

What about odd $N$?  We already know that the theory is anomalous if $N$ is odd, but in fact the details depend again on the value of $N$ mod 8.    First of all, generalizing part of the discussion
in section \ref{threety}, one aspect of the anomaly for any odd $N$ is that in an irreducible representation of the canonical anticommutation relations $\{\lambda_k,\lambda_l\}=
2\delta_{kl}$, the operator $(-1)^\fF$ does not act.   Such an irreducible representation has dimension $2^{(N-1)/2}$.   To construct such a representation,
we can use the previous definition of $N-1$ gamma matrices $\gamma_1,\cdots,\gamma_{N-1}$ acting on $(N-1)/2$ qubits, and then set\footnote{The sign in this formula
distinguishes the two irreducible representations of the odd $N$ Clifford algebra;  in what follows it does not matter
which sign we pick.} $\gamma_N=\pm\i^{(N-1)/2}\gamma_1
\gamma_2\cdots\gamma_{N-1}$.   In such an irreducible representation of the odd $N$ Clifford algebra, the product $\gamma_1\gamma_2\cdots\gamma_N$ is a $c$-number. There
can be no unitary operator $(-1)^\fF$ acting on this representation that conjugates $\gamma_k$ to $-\gamma_k$ for all $k$, since such an operator would reverse the sign of the $c$-number
$\gamma_1
\gamma_2\cdots\gamma_N$.

  \begin{table}
\begin{center}
\begin{tabular}{ |c|c|c| c|c |} 
 \hline
  $N~{\mathrm {mod}}~8 $&1 & 3&5&7 \\ \hline
 $\h{\tT}$ & $\tT$  & $\tT'$ &$\tT$&$\tT'$ \\ \hline
 $\h{\tT}^2$ & 1 & $-1$ & $-1$ &$1$ \\ 
 \hline
\end{tabular}
 \caption{\label{tabulk} This table summarizes the anomalies for odd $N$ mod 8.   The four cases are distinguished by whether the symmetry $\h\tT$ that is realized is
 $\tT$ or $\tT'=\tT(-1)^\fF$, along with the sign of $\h{\tT}^2$.}   
\end{center}
\end{table}

One can now ask whether the symmetries $\tT$ and $\tT (-1)^\fF$ can be realized in an irreducible representation of the anticommutation relations.   The answer is that, depending on the value of $N$ mod 8, one
or the other of these symmetries can be realized, but not both.    Looking for a time-reversal symmetry, we set
\be\label{welk}\h\tT = \star \gamma_1\gamma_3\gamma_5\cdots\gamma_N.\ee
It satisfies
\be\label{elk} \h\tT\lambda_k\h\tT^{-1} =\begin{cases} \lambda_k& ~~~~~N\cong 1,5 ~{\mathrm{mod}}~8 \cr
                                                                                     -\lambda_k & ~~~~~~N\cong 3,7~{\mathrm{mod}}~8.\end{cases} \ee
 Thus, the symmetry that can be implemented is $\tT$ if $N$ is congruent to 1 or 5 mod 8,
or  $\tT(-1)^\fF$  if $N$ is congruent to 3 or 7 mod 8.    
There is also an anomaly in the sign of $\h\tT$:
\be\label{pelk}\h\tT^2=\begin{cases} 1& ~~~~~N\cong 1,7~{\mathrm{mod}}~8 \cr
                                                          -1 & ~~~~~N\cong 3,5~{\mathrm{mod}}~8.\end{cases}\ee
 So the four cases of odd $N$ mod 8 are distinguished by which symmetry is realized, and whether its square is 1 or $-1$, as summarized in Table \ref{tabulk}.

\subsection{Boundary Conditions and The Anomaly}\label{bctime}

We now want to consider open strings and to introduce time-reversal invariant boundary conditions.   The massless Dirac equation, on a flat worldsheet for simplicity, reads
$\slashed{D}\psi=0$ with
\be\label{diracc} \slashed{D}=\gamma^\tau \partial_\tau+\gamma^\sigma\partial_\sigma.\ee
In Euclidean signature, $\{\gamma_\mu,\gamma_\nu\}=2\delta_{\mu\nu}$; continuing this to a Lorentz signature worldsheet, 
we take $\gamma_\tau^2=-1$, $\gamma_\sigma^2=+1$.    Just to get a time-reversal symmetry of the massless Dirac equation, there are essentially two
options: the operations $\tT_1$ and $\tT_2$ defined by  $\tT_1\psi(\sigma,\tau)\tT_1^{-1}=\gamma_\tau \psi(\sigma,-\tau)$, $\tT_2\psi(\sigma,\tau)\tT_2^{-1}=\gamma_\sigma\psi(\sigma,-\tau)$
 are both symmetries of the massless Dirac equation.
Note that $\tT_1{}^2=-1$, $\tT_2{}^2=+1$.   However, in a theory that contains open strings, we need a time-reversal symmetry that is consistent with the usual
open-string boundary conditions of eqn. (\ref{twob}), which in the present context read
\be\label{urgh}\bigl.\gamma_\sigma\psi\bigr|=\veps\bigl.\psi\bigr|,~~~~~~~\veps=\pm 1. \ee
The symmetry that satisfies this condition is $\tT=\tT_2$, with $\tT^2=1$.   That is actually the reason that in 
section \ref{time}  we considered boundary theories that have a time-reversal symmetry with $\tT^2=1$ acting on fermions.     Similar boundary theories with a time-reversal symmetry
that satisfies $\tT^2=-1$ on fermions also exist,\footnote{A minimal realization involves two Majorana fermions $\lambda_1,\lambda_2$, with $\tT\lambda_1(\tau)\tT^{-1}=\lambda_2(-\tau)$,
$\tT\lambda_2(\tau)\tT^{-1}=-\lambda_1(\tau)$.} but are not relevant if the bulk theory has a time-reversal symmetry with $\tT^2=1$.

Thus we take the bulk time-reversal symmetry to act by
\be\label{bulksym}\tT\psi(\sigma,\tau)\tT^{-1}=\gamma_\sigma \psi(\sigma,-\tau). \ee
As in section \ref{majotime}, the choice of sign on the right hand side of this relation does not matter.   Replacing
 the right hand side of eqn. (\ref{bulksym}) by $-\gamma_\sigma \psi(\sigma,-\tau)$ would amount to replacing $\tT$ with $\tT (-1)^\fF$.  This would amount to just using a different set of generators for the constraints that are going to be imposed.
 In Euclidean signature, with a boundary at $\sigma=0$, the reflection symmetry $\rR$ that is related to $\tT$ by $\sf{CPT}$ will transform a boundary fermion by
 \be\label{nsym}\rR(\psi(0,\tau)) =\gamma_\sigma \psi(0,-\tau).\ee  Here we are generalizing eqn. (\ref{belko}) to relate the action of $\tT$ and $\rR$ on arbitrary boundary operators.
 
 Type I superstring theory contains D9-branes.  When a string  ends on a D9-brane at $\sigma=0$, all worldsheet fermions satisfy the same boundary condition, which we can take
 to be
 \be\label{ulksym} \bigl.\psi\bigr| = \bigl.\gamma_\sigma\psi\bigr|.\ee
 A minus sign on the right hand side would be inessential, as it could be removed by a discrete chiral rotation $\psi\to \h\gamma\psi$, $\h\gamma=\gamma_\sigma\gamma_\tau$.
 This would reverse the sign of $\tT$, but as we have just explained, that sign is also inessential.   
 
 Now let us discuss Dp-branes of Type I superstring theory.   To discuss a Dp-brane, we take $p+1$ worldsheet bosons $X^k$, $k=0,\dots,p$ to satisfy Neumann boundary
 conditions, while the other $9-p$, which we will call $Y^r$, $r=1,\cdots 9-p$, satisfy Dirichlet boundary conditions.  The $X$'s parametrize the worldvolume of the brane, and the $Y$'s
 parametrize the normal directions.
 Worldsheet supersymmetry then tells us that the boundary conditions for the superpartners $\psi_X$ of the $X$'s should be unchanged, but we should flip the sign of
 the boundary condition for the superpartners $\psi_Y$ of the $Y$'s.   This flipped boundary condition is
  \be\label{nulksym} \bigl.\psi_Y\bigr| =- \bigl.\gamma_\sigma\psi_Y\bigr|.\ee
  
  From our study of Type II superstring theory, we know that this will produce an anomaly if $9-p$ is odd, and we can anticipate encountering a more refined anomaly
  when time-reversal symmetry is taken into account.     One simple way to analyze the problem is to return to the annulus of fig. \ref{two}.
  We think of the short direction from left to right in the annulus as a spatial direction, parametrized by $\sigma$, say with $0\leq \sigma\leq \pi$,  and the long direction as the (Euclidean or Lorentzian) time
  direction, parametrized by $\tau$, with $\tau\cong \tau+L$, $L\gg\pi$.  Let $\psi$ be the superpartner of one of the $X$'s or $Y$'s.   Depending on the boundary conditions satisfied by $\psi$ at $\sigma=0,\pi$ and on the spin structure,
  the field $\psi$ may or may not have a zero-mode in the $\sigma$ direction; if such a mode is present, it will propagate in the $\tau$ direction as a massless Majorana fermion $\lambda$.
  This mode will transform under $\tT$ with some sign, $\tT\lambda(\tau)\tT^{-1}=\veps \lambda(-\tau)$ with $\veps=\pm 1$.   Given that we have defined $\tT$ microscopically with
  a $+$ sign in eqn. (\ref{bulksym}), the mode $\lambda$, if it exists, will have $\veps=+1$ if the boundary condition satisfied by $\psi$ at $\sigma=0$ has a $+$ sign as in eqn.
  (\ref{ulksym}), but $\veps=-1$ if $\psi $ satisfies the opposite boundary condition (\ref{nulksym}).   This means that if we flip the sign in the boundary condition at $\sigma=0$
  from $+$ to $-$, we either remove a zero-mode $\lambda$ with $\veps=1$, or we add a zero-mode $\lambda$ with $\veps=-1$.   Either way, the coefficient of the mod 8
  anomaly that we studied in section \ref{majotime} is shifted by $-1$.   
  
  In the case of a Dp-brane, the boundary condition is flipped for $9-p$ worldsheet fermions, and this gives an anomaly coefficient $-(9-p)$.   The most obvious way to cancel the anomaly
  is to add $9-p$ boundary fermions that transform under time-reversal with $\veps=+1$.   This gives one way to construct a Type I Dp-brane, for any $p$.
  
  However, there are  alternatives.   First of all, since the anomaly is a mod 8 effect, boundary fermions with $\veps=+1$ (or $-1$) can be freely added or removed in groups of 8.
  Second, since a pair of boundary fermions with opposite signs of $\veps$ make an anomaly-free system, such pairs can also be freely added or removed.
  
  Combining these effects, we see that $n$ boundary fermions with $\veps=+1$ can be replaced with $8-n$ boundary fermions with $\veps=-1$.   This means that for any $p$,
  it is possible to make a Type I Dp-brane using at most 4 boundary fermions.
  
  But in fact, there is another way to reduce the required number of boundary fermions.   There is one nonzero value of the anomaly coefficient at which no boundary fermions
  at all are needed, and this is 4.   To see this, recall that as indicated in Table \ref{tabull}, for $N=4$ mod 8, there is no anomaly involving $(-1)^\fF$; in particular,
  this operator can be defined so that it squares to 1 and commutes with $\tT$, as it is supposed to.   The anomaly for $N=4$ mod 8  consists solely in the fact that $\tT^2=-1$,
  rather than $+1$.   But as we analyzed in section \ref{time}, a purely bosonic boundary theory with symplectic Chan-Paton factors can have $\tT^2=-1$.  
  Thus an example of a system with the $N=4$ anomaly is a two-dimensional boundary Hilbert space $\H_0$ on which $\tT$ acts as in eqn. (\ref{nff}) while
  $(-1)^\fF=1 $  (or $-1$).   
     Actually, this realization of the anomaly can be obtained
  from a fermion system with $N=4$ by projecting onto the subspace of the fermion Hilbert space with $(-1)^\fF=1$ (or $-1$).   Four Majorana fermions can be quantized in a four-dimensional
  Hilbert space, and the subspace with $(-1)^\fF=1$ (or $-1$) is two-dimensional; in an appropriate basis, $\tT$ acts on that space as in (\ref{nff}).

  With these facts in mind, we see that a Type I D5-brane can be constructed with no boundary fermions at all, and instead with a Chan-Paton symmetry group of symplectic type.
  This in fact gives the standard supersymmetric D5-brane or $\overline{\mathrm{D5}}$-brane of Type I superstring theory, depending on the choice of sign\footnote{\label{reversing} Reversing
  the sign of the $(-1)^\fF$ operator will give a minus sign for every Ramond boundary labeled by the brane.  As explained in section \ref{antibranes}, including this factor
  is the operation that in general  converts a supersymmetric brane into an antibrane.}
  of $(-1)^\fF$.
 Indeed, these branes were originally shown to carry symplectic Chan-Paton factors based on considerations somewhat similar to what 
 has just been explained \cite{Small}.
  
  We can now see that for any $p$, it is possible to construct a Type I Dp-brane with at most two boundary fermions.    In the case of a D4-brane, the anomaly coefficient because
  of the fermion boundary conditions is $-5$; we can cancel this with a single boundary fermion with $\veps=1$, contributing 1 to the anomaly, and the bosonic factor $\H_0$,
  contributing 4.   For a D6-brane, the anomaly coefficient because of the fermion boundary conditions is $-3$; this can be canceled with a single boundary fermion with $\veps=-1$,
  contributing $-1$ to the anomaly, and the bosonic factor $\H_0$, again contributing 4.    For Dp-branes with other values of $p$, the anomaly can be canceled with just
  one or two boundary fermions with $\veps=1$ or $\veps=-1$.

  \subsection{Type I D-Branes}\label{propbranes}
  
  \subsubsection{General Properties}\label{genprop}
  
 A physical state of Type I superstring theory is required to be invariant under the GSO projection, which is generated by $(-1)^\fF$, and under the orientifold projection, generated by $\rR$.
For operators inserted on a given brane boundary, both conditions are conveniently studied by looking at vertex operators: an allowed vertex operator should be invariant both under $(-1)^\fF$ and under
a reflection $\rR$ that leaves fixed the point at which the vertex operator is inserted, as in fig. \ref{five}(b).  

We can identify several questions of qualitative interest concerning a D-brane, and make some preliminary remarks about the vertex operators
relevant to these questions.

 (1)  Is a given D-brane stable or does it support a tachyonic mode?  Tachyon vertex operators of the Dp-Dp system, for any $p$, can be described as follows.
 (To decide if a given Dp-brane is stable, one also has to look at the Dp-D9 system.)
 Ignoring possible boundary fermions and Chan-Paton factors, the  open-string
 vertex operator for a tachyon state of momentum $k$ along a D-brane worldvolume is $V_\tach=k\cdot \psi e^{\i k\cdot X}$, where $X^m$ are coordinates along the $D$-brane
 worldvolume and $\psi^m$ are the superpartners of $X^m$.   This vertex operator is odd under $(-1)^\fF$ but, given the sign choices in eqns. (\ref{nsym}) and
 (\ref{ulksym}), it is even under  $\rR$.   To make a physical tachyon operator, we must combine $V_\tach$ with boundary fermions and Chan-Paton
 factors in such a way as a make an operator invariant under both $(-1)^\fF$ and $\rR$.
 
 (2)  Does a D-brane support gauge fields?   The vertex operator for a massless gauge field on a D-brane, ignoring Chan-Paton factors and boundary fermions,
was introduced in eqn. (\ref{newg}).   
 This operator is even under $(-1)^\fF$, but is odd under $\rR$, as previously noted.
 
 (3) Every D-brane is capable of transverse motion -- fluctuations in the normal direction to the brane worldvolume.   A more subtle question is whether a brane is
 reducible -- is it a superposition of separate components that can be displaced from each other in the transverse directions?   In \cite{Sen1,Sen2,Sen3}, certain
 nonsupersymmetric D-branes were shown to be irreducible.   We will generalize this analysis, showing that some branes are irreducible while
 some plausible looking constructions lead to branes that
 actually are reducible. 
  The vertex operator $V_\tr$ that describes transverse motion of a brane (if Chan-Paton factors and boundary fermions are ignored) is similar to $V_\gauge$,
 except that $\partial_\tau X$ is replaced by $\partial_\sigma Y$, the normal derivative of the worldsheet coordinates $Y$ that describe motion normal to the brane. $V_\tr$ is
 invariant under $(-1)^\fF$ just like $V_\gauge$, but 
 since the normal derivative is invariant under a reflection of the boundary, it is also invariant under $\rR$.  The operator $V_\tr $ is $X$-independent and therefore describes
 overall motion of a brane in the normal directions.  A signal of reducibility of a brane is that it admits another $X$-independent vertex operator that, like $V_\tr$, is invariant under
 both projections.   
 
 The behavior of $V_\tach, \,V_\gauge,$ and $V_\tr$ under the two projections is summarized in Table \ref{tabul}.   
 We note that this table does not contain any operators that are odd under both $(-1)^\fF$ and $\rR$, though certainly such vertex
 operators exist.
 
  \begin{table}
\begin{center}
\begin{tabular}{ |c|c|c| c|} 
 \hline
  & $V_\tach$ & $V_\gauge$ &$V_\tr$ \\ \hline
 $(-1)^{\fF}$ & $-1$ & 1 &1 \\ \hline
 $\rR$ & 1 & $-1$ & 1 \\ 
 \hline
\end{tabular}
 \caption{\label{tabul} This table summarizes the transformation of $V_\tach$, $V_\gauge$, and $V_\tr$ under $(-1)^{\fF}$ and $\rR$, ignoring boundary fermions and Chan-Paton
 factors.}   
\end{center}
\end{table}

 It is now relatively straightforward for the various D-branes of Type I superstring theory to answer the three questions just posed as well as others that we will discuss along the way.
 
 \subsubsection{D9-Branes and D1-Branes}\label{d9d1}
 
 For D9-branes and D1-branes, the anomaly coefficient vanishes, so there is little to say.   The usual supersymmetric D9-branes and D1-branes are entirely anomaly-free.
 Perhaps the only point worth mentioning is that these branes must be quantized with Chan-Paton factors of orthogonal, not symplectic, type, since symplectic Chan-Paton
 factors would contribute an anomaly coefficient of 4, as explained in section  (\ref{bctime}).
 
 \subsubsection{D8-Branes and D0-Branes}\label{d8d0}
 
 D8-branes and D0-branes have an anomaly coefficient of $-1$ mod 8, which can be canceled by adding a single boundary fermion $\lambda$ that transforms under $\tT$ with
 $\veps=+1$.   As in the discussion of  nonsupersymmetric Type II  D-branes in section \ref{branestwo}, the D8-D8  and D0-D0 systems have two types of vertex operator, schematically of the form 
 $V$ or $\lambda V$, where $V$ is constructed from the usual worldsheet fields.  Looking at Table \ref{tabul}, we see that a vertex operator of type $V$ can describe
 transverse motion of the brane, but not gauge fields or tachyonic modes along the brane, as the relevant vertex operators are odd under either $(-1)^\fF$ or $\rR$.  
 
 If we consider a system of $k$ D0- or D8-branes, we can make vertex operators $\a V_\gauge$, where $\a$ is a $k\times k$ antisymmetric matrix, a generator of $\SO(k)$.
 In other words, D0-branes and D8-branes have real Chan-Paton factors, along with the boundary fermion $\lambda$.   An important detail, which will play an important role
 shortly, is that, as always with real Chan-Paton factors, the gauge group with $k$ D0- or D8-branes is $\oO(k)$, just $\SO(k)$.    This means that for a single D0- or D8-brane,
 the gauge group is not quite trivial; rather, it is $\oO(1)\cong \Z_2$.   
  
 What about tachyons?   In order for $\lambda V$ to be even under $(-1)^\fF$, $V$ must be odd, like $V_\tach$.   However, if $V$ is odd under $(-1)^\fF$
 (and thus fermionic), then $\lambda V$ transforms oppositely to $V$ under $\rR$, since $\rR$ reverses the order with which $\lambda$ and $V$ are inserted on the worldsheet boundary.
 As $V_\tach$ is even under $\rR$, $\lambda V_\tach$ is odd.
 
 It turns out that the D0-D9 system is also tachyon-free so the D0-brane is entirely stable.  (By contrast,\footnote{This was pointed out by A. Sen.}
  the D8-D9 system is tachyonic, so the D8-brane is not entirely stable.)  
 This is an important result of \cite{Sen1,Sen2,Sen3}:   Type I superstring theory has D0-branes  that are stable, though nonsupersymmetric, because the tachyon
 is removed by the orientifold projection.   
 
It was shown in  those same papers  that the states of the D0-D9 system transform in spinorial representations  of the gauge group of the Type I theory.
This follows straightforwardly upon quantizing the D0-D9 strings in the Ramond sector.   As always for a free worldsheet theory, 
 the ground state energy in the Ramond sector of the D0-D9 system
 vanishes.  Because of the opposite boundary conditions at the two ends of the string, the transverse bosonic oscillators $Y^i$ of the D0-D9 system have no zero-modes; in the
 Ramond sector, by worldsheet supersymmetry, the same is true of their superpartners $\psi_Y^i$.   The fields that do have zero-modes are a bosonic oscillator $X^0$ that parametrizes the ``time'' 
 direction, that is, the worldline of the D0-brane, and its superpartner $\psi_X^0$.  There is also a Majorana fermion $\lambda$ that propagators on the D0 boundary.
 Quantizing $\psi_X^0$ and $\lambda$ gives two states, one of which survives the GSO projection.  This state is a fermion because we are in the Ramond sector.\footnote{It has
 half-integer spin because of fermion zero-modes of the D0-D0 system in the Ramond sector.   These modes are $\SO(32)$-invariant so do not affect the discussion
 of how the states of this system transform under $\SO(32)$.}   Because the
 orientifold projection identifies D0-D9 and D9-D0 strings, this state is a Majorana fermion.     Taking into account
 the Chan-Paton labels of the D9-brane system, the D0-D9 system actually has 32  such Majorana modes $\chi_i$, $i=1,\cdots,32$, transforming in the vector representation of $\SO(32)$.
 These modes can propagate along the D0-brane world-volume, which is parametrized by $t=X^0$.
An effective action for these modes is
 \be\label{effac}\frac{\i}{2}\sum_{i=1}^{32}\int \d t \chi_i\frac{\d}{\d t}\chi_i . \ee
 Quantization of those modes gives $2^{32/2}=2^{16}$ states that form a  spinor representation of $\Spin(32)$, the double cover of $\SO(32)$.   However, we have to remember that the
 D0-brane actually carries a $\Z_2$ gauge symmetry, under which the states of the D0-D9 system are odd.   That means in particular that the $\chi_i$ are odd under $\Z_2$.
So the $\Z_2$ generator is the chirality operator $\bar\chi=\chi_1\chi_2\cdots\chi_{32}$, which actually is an element of the center of $\Spin(32)$.
   Because this $\Z_2$ symmetry is a gauge symmetry, the $2^{16}$ states
obtained by quantizing the $\chi_i$ have to be projected to $\bar\chi$ invariant states, and so provide an irreducible representation of $\Spin(32)/\Z_2$ where here $\Z_2$ is the subgroup
of the center of $\Spin(32)$ that 
is generated by $\bar\chi$.

While, as just described, the  ground states in the Ramond sector of the D0-D9 system are in a spinor representation of $\Spin(32)/\Z_2$ (a representation that can be constructed
by quantizing 32 gamma matrices), excited states  are in spinorial representations of the same group 
(arbitrary representations 
that are odd under a $2\pi$ rotation in $\SO(32)$), since the D9-D9 vertex operators  that map between these states  are in representations of  $\SO(32)/\Z_2$ (and the D0-D0
vertex operators are $\SO(32)$-invariant).   
Likewise the Neveu-Schwarz sector states of the D0-D9 system transform in spinorial representations, because of fermion zero-modes that appear in the quantization.
 This is an inevitable consequence of the result in the Ramond sector, since there are $\SO(32)$-invariant vertex operators (for example, Ramond sector vertex operators of the
 D0-brane) that exchange the two sectors.  For the same reason, both sectors are invariant under the same $\Z_2$ subgroup of $\Spin(32)$ and transform as representations of
 $\Spin(32)/\Z_2$.
 
 The significance of all this is to make possible duality between Type I superstring theory and the heterotic string.
  In string perturbation theory, Type I superstring theory has a potential  $\OO(32)$
 gauge group, coming from modes of the D9-D9 system.   But duality with the heterotic string predicts that the gauge group is really $\Spin(32)/\Z_2$.   
 The analysis of the D0-D9 spectrum both extends and reduces the naive $\OO(32)$ symmetry: it is reduced because the disconnected component of $\OO(32)$
 is explicitly violated by the $\bar\chi$ projection, but extended to a  double cover of the gauge group because the D0-D9 states are in spinorial representations.
This extension and reduction of the 
 Type I gauge symmetry  make possible heterotic-Type I duality.   That was one of the main insights of \cite{Sen1,Sen2,Sen3}.

 \subsubsection{D7-Branes and D$(-1)$-Branes}\label{d7d(-1)}
 
 D7-branes and D$(-1)$-branes have an anomaly coefficient of $-2$, which can be canceled by adding two boundary fermions $\lambda_1,\lambda_2$, each $\tT$-even.
  
 The worldvolume of a D$(-1)$-brane is a point, with no tangential coordinates $X$, so a D$(-1)$-brane does not support tachyon vertex operators $V_\tach \sim k\cdot \psi e^{\i k\cdot X}$
 or gauge vertex operators $V_\gauge\sim \partial_\tau X$.   But we can ask whether the D$(-1)$-brane is reducible.   For the D7-brane, all three questions involving
 stability, gauge symmetry, and reducibility are applicable.   
 
 There are three types of vertex operators to consider, schematically of the form $V$, $\lambda_i V$ with $i=1,2$, and $\lambda_1\lambda_2 V$.  
 As usual, 
 vertex operators of the form $V$ describe normal motion of the brane but not gauge fields or tachyons.  For a vertex operator $\lambda_i V$ to be invariant under both $(-1)^\fF$ and $\rR$,
 $V$ must be odd under both of those symmetries.  (The fact that $V$ must be odd under $\rR$ was explained in section \ref{d8d0}.) 
 Reviewing Table \ref{tabul}, we see that vertex operators $\lambda_i V$ do not describe either gauge fields or
 tachyonic modes or normal motion.    In order for $\lambda_1\lambda_2 V$ to be invariant under both projections, $V$
 must be even under $(-1)^\fF$ and odd under $\rR$.   From the table, we see that $V_\tach$ does not have the relevant properties, so the D7-D7 system is tachyon-free,\footnote{However,
 the D7-D9 system is tachyonic, so the D7-brane is not entirely stable.}
  and $V_\tr$ does not have the relevant properties, so the nonsupersymmetric D$(-1)$-brane and D7-brane are irreducible.   
 However,
 $V_\gauge $ does have  the necessary properties.   
So a nonsupersymmetric D7-brane supports a $\UU(1)$ gauge field, with 
  the vertex operator $\lambda_1\lambda_2 V_\gauge$.   The  charge generator is ${\sf Q}=\i\lambda_1\lambda_2$.  This  is a conserved charge in the boundary theory
and plays the role that in open-string theory is usually
 played by a Chan-Paton matrix.   For future reference, note that ${\sf Q}$ coincides (up to a possible choice of sign) 
 with the operator $(-1)^\fF$ in the boundary Hilbert space that anticommutes with $\lambda_1,\lambda_2$.
 
Of course, there are more obvious ways to get $\UU(1)$ gauge symmetry on a brane.   For example, a pair of branes with real Chan-Paton factors
have $\SO(2)\cong \UU(1)$ gauge symmetry.   The construction  was reviewed in section \ref{time}. This brane is reducible; it can separate into its two components.  The reducibility shows
up in the existence of $X$-independent vertex operators of the form $V_\tr \otimes U$, where $U$ is a symmetric $2\times 2$ matrix that acts on the Chan-Paton labels.  From 
eqn. (\ref{rff}), it follows that such a vertex operator is $\rR$-invariant.
For $U$ a multiple of the identity, this vertex operator describes overall transverse motion of the pair of branes; for more general $U$, it describes a transverse motion in which the brane separates
into its two components.   The nonsupersymmetric D7-brane and D$(-1)$-brane have no analog of the vertex operators $V_\tr \otimes U$ with $U$ not a multiple of the identity; they are
 irreducible.

We can see the difference between the nonsupersymmetric D7-brane and a more obvious construction that leads to the same gauge symmetry by considering the D7-D9 system
in the Ramond sector.   Quantization of $\lambda_1$ and $\lambda_2$ leads to a pair of states, with charges $\pm 1$ under the gauge symmetry of the D7-brane (assuming
that the charge generator is normalized as ${\sf Q}=\i\lambda_1\lambda_2=(-1)^\fF$).   Up to this point, we would see the same spectrum if we more naively assume
that the D7-brane gets its gauge symmetry via 
$\SO(2)$ Chan-Paton factors, which also  imply a doubling of the spectrum.  However, the relation ${\sf Q}=(-1)^\fF$  implies that among the two states
obtained by quantizing $\lambda_1$ and $\lambda_2$, the state of ${\sf Q}=1$ has $(-1)^\fF=1$ and the state of ${\sf Q}=-1$ has $(-1)^\fF=-1$.  Now consider  the oscillator modes of
the D7-D9 system in
the Ramond sector.   In this sector, the fermion partners $\psi_X$ of the worldsheet fields $X$ that parametrize the brane worldvolume have zero-modes.  These zero-modes transform in the vector representation of $\SO(1,7)$ (the Lorentz group of the D7-brane worldvolume). As usual,
quantization of these modes produces massless states that are spinors of $\Spin(1,7)$,   On these states,  we must impose the GSO projection.   
Because the ${\sf Q}=1$ and ${\sf Q}=-1$ modes obtained
by quantizing the boundary fermions have opposite signs of $(-1)^\fF$, we must impose opposite GSO projections on states of ${\sf Q}=1$ and ${\sf Q}=-1$.
Thus we get, with the appropriate choice of orientation of the brane, left-handed spinors of $\Spin(1,7)$ with ${\sf Q}=1$ and right-handed spinors of $\Spin(1,7)$ with ${\sf Q}=-1$.

This result is precisely what is needed for consistency of the effective field theory.  As the two spinor representations of $\Spin(1,7)$ are complex conjugates of each other,
positive chirality spinors with ${\sf Q}=1$ must be accompanied by negative chirality spinors with ${\sf Q}=-1$.   The more naive picture with $\SO(2)$ Chan-Paton factors
would not lead to a consistent spectrum; in that picture, the fermion chirality would be independent of the charge.

The nonsupersymmetric D$(-1)$-brane plays a role in reconciling the gauge group of Type I superstring theory, which in perturbation theory appears to be potentially $\OO(32)$,
with the $\Spin(32)/\Z_2$ gauge symmetry of the heterotic string.   D$(-1)$-brane anplitudes explicitly violate the invariance under the disconnected component of $\OO(32)$.
 This was seen previously on the basis of the relationship between D-branes and K-theory \cite{dk},
and now we can see it via a direct  worldsheet construction.\footnote{Of course, we also saw in another way in section \ref{d8d0} that the disconnected component
is not a symmetry.}   The D$(-1)$ brane is not a particle or a physical object; it is better understood as an instanton, a localized
object that exists at a point in spacetime.   As always for instantons, an anomaly can arise if the measure for the fermion zero-modes in the instanton field is not
invariant under a classical symmetry.   (In fact, we encountered this phenomenon in introducing the anomaly in $(-1)^\fF$ in section \ref{threety}.)
In the present case, we can find the relevant fermion zero-modes by quantizing the D$(-1)$-D9 strings.
The ground state of the D$(-1)$-D9 system in the Ramond sector has zero energy, because of the usual cancellation between bosons and fermions in the Ramond sector,
and it is fermionic, as we are in the Ramond sector.   The  bosonic fields $X^I$ and  their superpartners $\psi^I$ all satisfy opposite boundary conditions at the two ends of a
D$(-1)$-D9 string, so none of them have zero-modes.   Quantizing the two boundary fermions $\lambda_1,\lambda_2$ at the D$(-1)$-brane end of the string gives a pair of
states, of which one survives the GSO projection.   Taking into account the $\SO(32)$ Chan-Paton factors at the other end of the string, this gives us a set of 32 fermionic
ground state modes, transforming in the vector representation of $\SO(32)$.  Let us call these modes $\eta_1,\eta_2,\cdots,\eta_{32}$.   They live at the point in spacetime
where the D$(-1)$-brane lives, and do not propagate anywhere.  Instead, they are analogous to what in field theory would be fermion zero-modes in an instanton field.   
  The measure $\mu =\d \eta_1\d \eta_2\cdots \d\eta_{32}$ for integration over  these modes is invariant under $\SO(32)$ but not under the disconnected component of
$\OO(32)$.  Hence invariance under that disconnected component is lost.

 To complete the picture, we need to know that all Ramond and Neveu-Schwarz sectors of
all Dp-D9 sytems for all $p$ are invariant under the same $\Z_2$ subgroup of $\Spin(32)$, so that the gauge group is really $\Spin(32)/\Z_2$.
 This is true because there are $\SO(32)$-invariant vertex operators
that map between all of these sectors.   For example, for any $p,p'<9$, vertex operators of the Dp-Dp$'$ system are $\SO(32)$-invariant and map the Dp-D9 system to the Dp$'$-D9 system;
similarly, for each $p$, SO(32)-invariant Ramond vertex operators exchange the Ramond and Neveu-Schwarz sectors.

\subsubsection{D2-Branes}\label{d2}

The D2-brane is a sort of mirror of the D0-brane, which we have already analyzed, so we can analyze it briefly.   The anomaly coefficient of the D2-brane is $-7$, which is equivalent
to 1 mod 8.   So we can cancel the anomaly with a single $\tT$-odd boundary fermion, which we will call $\rho$.   (We will write generically $\lambda$ and $\rho$ for $\tT$-even
and $\tT$-odd boundary fermions, respectively.)   

We have to consider vertex operators of the general type $V$ and $\rho V$.   As usual, vertex operators $V$ describe the overall center of mass motion of a brane, but not
tachyons or gauge symmetry.   For a vertex operator $\rho V$ to commute with the GSO projection, $V$ must be odd under $(-1)^\fF$; in that case, for $\rho V$ to commute with $\rR$,
$V$ itself must be $\rR$-invariant.  The last statement holds because if $V$ is GSO-odd, then it is fermionic; then this being so, when we apply $\rR$ to $\rho V$, we get a minus
sign because $\rho$ is odd under $\rR$, and another minus sign from reordering $\rho$ and $V$.   So $\rho V$ is $\rR$-invariant if and only if $V$ itself is $\rR$-invariant.

Given this, a look back to Table \ref{tabul} shows that the D2-brane has a  vertex operator $\rho V$ with $V=V_\tach$, but not with $V=V_\gauge$ or $V=V_\tr$.
Thus the D2-brane is tachyonic, but it is irreducible and has no gauge symmetry.

\subsubsection{D3-Branes}\label{d3}

Similarly, the D3-brane is a sort of mirror of the D$(-1)$-brane.  The anomaly coefficient is $-6$, or equivalently 2 mod 8.  So we can cancel the anomaly with a pair of $\tT$-odd boundary
fermions $\rho_1$ and $\rho_2$.   

As in the D2-brane case, there are tachyon vertex operators of the form $\rho_i V_\tach$.   These are actually charged under a $\UU(1)$ gauge symmetry that is associated to the vertex
operator $\rho_1\rho_2 V_\gauge$.   This nonsupersymmetric D3-brane is irreducible, since $\rho_1\rho_2 V_\tr$ is odd under $\rR$.

The analysis of the D3-D9 system in the Ramond sector leads to considerations similar to what we explained for the D7-D9 system, since the two spinor representations of
$\Spin(1,3)$ are complex conjugates.

\subsubsection{D4-Branes}\label{d4}

For the D4-brane, the anomaly coefficient is $-5$, or 3 mod 8, so we can cancel it with three $\tT$-odd boundary fermions $\rho_1,\rho_2,\rho_3$.

The gauge group is now $\SO(3)$, or more precisely $\SU(2)$ as we explain momentarily, with vertex operators $\frac{\i}{2}\epsilon_{ijk}\rho^j\rho^k V_\gauge$ 
($\epsilon_{ijk}$ is the antisymmetric tensor with $\epsilon_{123}=1$).
The boundary conserved charges that play the role usually played by Chan-Paton matrices are ${\sf Q}_k=\frac{\i}{2}\epsilon_{ijk}\rho^j\rho^k $.   There are tachyon
vertex operators $\rho_i V_\tach$, transforming as spin 1 under the gauge group.   

For a vertex operator $\rho_1\rho_2\rho_3 V$ to be invariant under both projections, $V$ itself must be odd under both $(-1)^\fF$ and $\rR$.   None of the vertex operators
in Table \ref{tabul} has this property, so operators of this kind are not associated to any obvious qualitative properties of the D5-brane.

To see that in Type I superstring theory, the D4-brane gauge group is really $\SU(2)$ rather than $\SO(3))$, we consider the D4-D9 system.   To quantize the D4-D9 strings,
we have to quantize in particular the three Majorana fermions $\rho_i$ that live on the D4 boundary.  Quantizing those modes gives states transforming in the
two-dimensional representation of $\SU(2)$, so the whole D4-D9 Hilbert space transforms in that representation.

The same Type I D4-brane has another construction that actually is equivalent.   Here, we cancel the anomaly by $\SU(2)$ Chan-Paton factors, shifting the anomaly
by 4 as explained in section \ref{bctime}, and also adding a single $\tT$-even boundary fermion $\lambda$ contributing $+1$ to the anomaly.   The $\SU(2)$ Chan-Paton factors
are  represented by a two-dimensional Hilbert space $\H_0$ on which $(-1)^\fF$ acts trivially (or as multiplication by $-1$) and $\tT$ acts as in eqn. (\ref{nff}).

We will first give a few simple indications that the two constructions are equivalent, and then  a general proof.   First we look at the brane tensions.
As in our discussion of nonsupersymmetric Type II D-branes in section \ref{branestwo}, a factor in the brane tension is the partition function of the boundary degrees of freedom.
In the construction based on the three $\tT$-odd fermions $\rho_i$, the boundary partition function is $(\sqrt 2)^3=2\sqrt 2$, with one factor of $\sqrt 2$ for each boundary fermion $\rho_i$.
In the description with the Chan-Paton factors, the boundary partition function is again $2\sqrt 2$, with a factor of 2 from the Chan-Paton factors and a factor $\sqrt 2$ from $\lambda$.
In each description, the gauge group is $\SU(2)$.   In the description with the fields $\rho_i$, we have already found tachyon vertex operators in the spin 1 representation of 
the gauge group.   In the other description, the $\rR$-invariant and GSO-invariant tachyon vertex operators are $\i \vec\sigma \lambda V_\tach$, where $\vec\sigma$ are the Pauli 
matrices acting on $\H_0$.  We showed in section \ref{time} that $\vec\sigma$, as an operator on $\H_0$, is odd under $\rR$;  $\lambda V_\tach$ is also odd.

For a complete proof, we simply compare the operator algebras of the two boundary states. 
Here is is convenient to use a Hamiltonian formulation and to work with $\tT$ symmetry rather than $\rR$ symmetry.
   Whether described by the $\rho_i$ or by $\H_0$ together with $\lambda$,
the boundary degrees of freedom of the D4-brane do not have a completely satisfactory quantization; this boundary theory has an anomaly coefficient of 5.
However, the two boundary theories
 do have completely satisfactory operator algebras, and to show that they  are equivalent, it suffices to show that the two
operator algebras are the same, taking into account the $(-1)^\fF$ and $\tT$ symmetries.   If this is the case, then the same vertex operators  can be constructed in either
description.   It is not necessary to explicitly discuss gauge symmetries, since if the algebras of operators that can be used to construct vertex operators are the same in the two
cases, the gauge symmetries that we get when constructing vertex operators will inevitably also be the same.  

A convenient way to take the $\tT$ symmetry into account is to consider the algebra of $\tT$-invariant operators.   $\tT$-invariant operators do form an algebra, of course.
It is not necessary to explicitly discuss the $\tT$-odd operators, because a $\tT$-odd operator is $\i$ times a $\tT$-even operator, so if the algebras of $\tT$-invariant
operators match correctly, this match will persist when including operators that are not $\tT$-invariant.   (The ability to restrict to $\tT$-invariant operators and forget about $\tT$ does 
not have an obvious analog   in terms of $\rR$ symmetry; that
is why we express the argument in terms of $\tT$.)     We do have to keep track of the action
of $(-1)^\fF$ on the algebra of $\tT$-invariant operators.   In other words,  we have to consider the algebra of $\tT$-invariant operators as a $\Z_2$-graded algebra.   If two boundary
theories with  $(-1)^\fF$ and $\tT$ symmetries have the same $\Z_2$-graded algebras of $\tT$-invariant operators, then they are equivalent.

In the description by the $\rho_i$, the algebra of $\tT$-invariant operators is generated by bosonic operators $b_k=\i {\sf Q}_k=-\epsilon_{ijk}\rho^j\rho^k$, along
with $\alpha=\i \rho_1\rho_2\rho_3$.   The $b_k$ commute
with $(-1)^\fF$, and generate a quaternion algebra $b_1^2=b_2^2=b_3^2=-1, $ $b_1 b_2=-b_2 b_1=b_3$.  Meanwhile $\alpha$ is odd
under $(-1)^\fF$, commutes with $b_k$, and satisfies
$\alpha^2=1$.  The isomorphism with the other description goes by
\begin{align}\label{dotek}  b_k& \leftrightarrow -\i \sigma_k \cr \alpha & \leftrightarrow \lambda ,\end{align}
where $\sigma_k$ are the Pauli matrices acting on $\H_0$.   It is straightforward to verify that this mapping preserves all algebraic relations as well as the $\Z_2$ grading.

\subsubsection{D6-Branes}\label{d6}

The D6-brane is a sort of mirror image of the D4-brane, with anomaly coefficient $-3$, or 5 mod 8.   We can cancel the anomaly by adding three $\tT$-even boundary fermions
$\lambda_1,\lambda_2,\lambda_3$.   

The gauge group is $\SO(3)$ or more precisely $\SU(2)$, as in the D4 case, with vertex operators $\frac{\i}{2}\epsilon_{ijk}\lambda^j\lambda^k V_\gauge$.   The D6-brane is tachyonic, just
like the D4-brane, but the details are different.   Vertex operators $\lambda_i V_\tach$ are odd under $\rR$, but $\lambda_1\lambda_2\lambda_3 V_\tach$ is even.
so the D6-brane is tachyonic, like the D4-brane, but the tachyon vertex operator is a gauge-singlet.

As in the D4-case, there is an equivalent description in which the boundary degrees of freedom are $\SU(2)$ Chan-Paton factors, contributing 4 to the anomaly, and a single
$\tT$-odd fermion $\rho$, contributing $-1$.   In this description, the gauge group is again $\SU(2)$, and there is a gauge singlet tachyon vertex operator $\rho V_\tach$.

The proof that the two descriptions are equivalent can again proceed by directly comparing the two boundary algebras.   In the description by the $\lambda$'s, the
algebra of $\tT$-invariant boundary operators  is generated by  $b_k=-\epsilon_{ijk}\lambda^j\lambda^k$, again generating a quaternion algebra, along with a $\Z_2$-odd operator
$\alpha=\lambda_1\lambda_2\lambda_3$,
now satisfying $\alpha^2=-1$.  The isomorphism (\ref{dotek}) is replaced by
\begin{align}\label{zotek}  b_k& \leftrightarrow -\i \sigma_k \cr \alpha & \leftrightarrow \i\rho .\end{align}

\subsubsection{D5-Branes}\label{d5}

The anomaly coefficient of the D5-brane is $-4$, or equivalently 4 mod 8.

As already explained in section \ref{bctime}, we can cancel this anomaly by endowing the D5-brane with a two-dimensional boundary Hilbert space $\H_0$ and
an $\SU(2)\cong \Sp(1)$ Chan-Paton gauge group.   In this way, we can make the usual supersymmetric D5-brane and $\overline{\mathrm{D5}}$-brane.   The two
differ by whether we consider the operator $(-1)^{\fF}$ to act as $+1$ or $-1$ on $\H_0$.  (The fact that this difference distinguishes branes and antibranes
was explained in footnote \ref{reversing}.)
 
But what if we cancel the anomaly by adding four $\tT$-even boundary fermions  $\lambda_1,\cdots,\lambda_4$?   
As a preliminary comment, it would be equivalent to cancel the anomaly by adding four $\tT$-odd boundary fermions
$\rho_1,\cdots, \rho_4$, since if the $\lambda_i$ are hermitian and $\tT$-even generators of a Clifford algebra, and we set $\bar\lambda=\lambda_1\lambda_2\lambda_3\lambda_4$,
then $\rho_i=\i \lambda_i\bar\lambda$ are hermitian and $\tT$-odd generators of an isomorphic Clifford algebra. This can also be inverted to define the $\rho$'s in terms of the
$\lambda$'s.  So the same boundary theory can be described in terms of either the $\rho$'s or the $\lambda$'s.   

If we cancel the anomaly by incorporating the $\lambda$'s, we get a brane that is actually reducible.   The reason is that in addition to the 
vertex operator $V_\tr$ that describes overall transverse motion of the brane, there is a vertex
operator $V_\tr{}'=\bar\lambda V_\tr$ that describes relative transverse motion of brane states with $\bar\lambda=1$
and $\bar\lambda=-1$.  

In the four-dimensional Hilbert space $\H$ obtained by quantizing the $\lambda_i$, let $\H_0$ be the two-dimensional subspace with $\bar\lambda=1$ and $\H_0'$ the 
two-dimensional subspace with
$\bar\lambda=-1$.   On each of these subspaces, $\tT^2=-1$, since in fact when we quantize four $\tT$-even Majorana fermions, $\tT^2=-1$ as an operator on $\H$.   
So endowing a D5-brane with the boundary Hilbert space $\H_0$ or $\H_0'$ is precisely the operation that leads to $\SU(2)$ Chan-Paton factors and to the standard
supersymmetric D5-brane and $\overline{{\mathrm D}5}$-brane.   

Intuitively, turning on the vertex operator $V_\tr$ displaces the part of the brane with $\bar\lambda=1$, which would appear to be a supersymmetric D5-brane,
from the part with $\bar\lambda=-1$, which would similarly appear to be a supersymmetric $\overline{{\mathrm D}5}$-brane.  So we are led to suspect
that the brane characterized by the four boundary fermions $\lambda_1,\cdots,\lambda_4$ is equivalent to a standard D5-$\overline{{\mathrm D}5}$ system.

We can check this by a slightly  more detailed computation of the spectrum of the brane with the boundary fermions.
This brane has vertex operators $V_{ij}=\frac{i}{2}[\lambda_i ,\lambda_j ]V_\gauge$, $i,j=1,\cdots,4$ that generate a gauge symmetry $\SO(4)\sim \SU(2)\times \SU(2)$.
In addition it has  tachyonic operators $W_i=\frac{1}{3!}\epsilon_{ijkl}\lambda^j\lambda^k\lambda^l V_\tach$, transforming as a bifundamental of $\SU(2)\times \SU(2)$.
All this matches the behavior of the standard D5-$\overline{{\mathrm D}5}$ system.

As in previous examples, the equivalence of the two descriptions can be found by comparing the two boundary algebras. 
In the description by the $\lambda$'s, the boundary algebra is a Clifford algebra ${\sf Cl}_4$  with four generators $\lambda_1,\cdots,\lambda_4$.  In the ``Chan-Paton''
description, the boundary algebra is the algebra $\A$ of linear transformations of the four-dimensional vector space $\H_0\oplus \H_0'$.
The Clifford algebra has an irreducible representation in a four-dimensional vector space with a two-dimensional subspace even under $(-1)^\fF=\bar\lambda$
and a two-dimensional odd subspace.  As a $\Z_2$-graded vector space, this representation space 
can be identified  with $\H_0\oplus \H_0'$.
That    gives an embedding of ${\sf Cl}_4$ in $\A$.  Counting dimensions,
${\sf Cl}_4$ has dimension $2^4=16$, and $\A$ has dimension $4^2=16$.   As these numbers coincide, the embedding of ${\sf Cl}_4$ in $\A$ is actually an isomorphism
between these two algebras, and therefore the two descriptions are equivalent.

\subsubsection{Reducible Branes}\label{morex}

For $p=1,5,9$, the usual supersymmetric D-branes and antibranes are irreducible.   For every other value of $p$, we have constructed an irreducible nonsupersymmetric Type I D-brane.
(These branes are equivalent to their own antibranes, for a reason explained at the end of section \ref{antibranes}.)    Various other constructions
of Dp-branes  present themselves, but it appears that these constructions give branes that are equivalent to a certain number of copies of the branes that
we have already constructed.   We will not attempt a general proof of this fact, but will illustrate it with an example.   

A D6-brane has an anomaly coefficient $-3$ mod 8, and we canceled the anomaly in section \ref{d6} by adding three $\tT$-even boundary fermions $\lambda_1,\cdots,\lambda_3$.
We could have equally well canceled the anomaly by adding five $\tT$-odd boundary fermions $\rho_1,\cdots, \rho_5$.

Every boundary fermion contributes a factor $\sqrt 2$ to the brane tension.  The second construction with the $\rho_i$ has 2 more boundary fermions, so it gives a brane whose tension is precisely
2 times the tension of the brane that we constructed in section \ref{d6} with the three $\lambda$'s.    This suggests that the brane constructed with the five $\rho$'s
might be equivalent to two copies of the brane constructed with the three $\lambda$'s.   

To describe two copies of a brane, we equip the boundary with a two-dimensional Chan-Paton state space $\H_b$.  
Since symplectic Chan-Paton factors would contribute to the anomaly, we need orthogonal Chan-Paton factors.   So we take $\tT=\star$ as an operator on $\H_b$.   Let $M_2(\C)$  be the algebra of $2\times 2$
complex matrices, and let $M_2(\C)[\lambda_1,\lambda_2,\lambda_3]$ be the algebra generated by $M_2(\C)$ together with the $\lambda$'s.  Similarly
let $\C[\rho_1,\rho_2,\cdots,\rho_5]$   be the algebra over $\C$ generated by the $\rho_i$.  To show that the two descriptions are the same, it suffices to show that,
as $\Z_2$-graded algebras, the $\tT$-invariant part of $M_2(\C)[\lambda_1,\lambda_2,\lambda_3]$ is isomorphic to the $\tT$-invariant part of $\C[\rho_1,\rho_2,\cdots,\rho_5]$.
Indeed, setting $\bar\lambda=\lambda_1\lambda_2\lambda_3$,  
 an isomorphism between the two takes the form
\begin{align}\label{dlme}   \i\sigma_2\lambda_i    & \leftrightarrow \i\rho_i,\hskip.3cm i=1,2,3\cr
                                        \sigma_1\bar\lambda &\leftrightarrow \i\rho_4\cr
                                           \sigma_3\bar\lambda&\leftrightarrow\i\rho_5.\end{align}

  \vskip1cm
 \noindent {\it {Acknowledgements}}    I thank O, Bergman, D. Freed, L. Rastelli, and A. Sen  for helpful comments.
  Research supported in part by NSF Grant PHY-2207584.

    \begin{appendix}
     
    \section{A Note on the $\sf{CPT}$ Theorem}\label{cpt}
    
  The following remarks on the $\sf{CPT}$ theorem might provide some context for the discussion in section \ref{time}.  The $\sf{CPT}$ theorem comes from the fact that a reflection of
two coordinates $(x_1,x_2,x_3,\dots,x_n)\to (-x_1,-x_2,x_3,\dots,x_n)$ is always a symmetry in Euclidean signature (since it is a $\pi$ rotation of the $x_1-x_2$ plane).   Wick rotating
to Lorentz signature
by $x_1\to \i x_1$, this operation becomes what one might call $\sf{RT}$, where $\rR$ reverses the sign of the space coordinate $x_2$, and $\tT$ reverses the sign of the time
coordinate $x_1$.  Therefore ${\sf{RT}}$ is unavoidably a symmetry of any relativistic quantum field theory.

 There are a few subtleties concerning the usual terminology for this theorem.
In the real world of three space dimensions, a relativistic theory
has  a parity symmetry
$\sf P$ that reverses the sign of all space coordinates  $(x_1,x_2,x_3)\to (-x_1,-x_2,-x_3)$  if and only if it has a reflection symmetry $\rR$ that reverses the sign of just one space coordinate
$(x_1,x_2,x_3)\to (-x_1,x_2,x_3)$, since the two differ
by a $\pi$ rotation of the $x_2-x_3$ plane.     
To state the theorem in a way that is uniformly valid in any dimension, even or odd, one might call it the $\sf{CRT}$ theorem, but in $3+1$ dimensions,
the theorem is traditionally called the $\sf{CPT}$ theorem.
Moreover, there is a subtlety concerning what is  $\tT$ and what is  $\sf{CT}$.   Traditionally, $\tT$ is defined to commute with conserved charges such 
as baryon and lepton number, which makes it possible for $\tT$ to be
 an approximate symmetry of ordinary matter.   Given this,  the continuation of a reflection symmetry from Euclidean signature to a time-reversal symmetry in Lorentz signature is
usaully called $\sf{CT}$, since it reverses the sign of conserved charges.
  That is built into the usual terminology
concerning the $\sf{CPT}$ theorem; if one reversed the definitions of $\tT$ and $\sf{CT}$, which might be natural purely from the point of view of relativistic field theory, one
would call it the $\sf{PT}$ theorem or possibly the $\sf{RT}$ theorem.   
The worldsheet symmetry that we call $\tT$ in this article might be called $\sf{CT}$ according to
standard conventions, since in the boundary theory that is introduced in section
\ref{majotime}, this symmetry reverses the sign of hermitian conserved charges such as $\i\lambda_1\lambda_2$.   However, it seems unduly clumsy  in the present context
to call this symmetry $\sf{CT}$, and we have chosen to call it  $\tT$.

\end{appendix}

 \bibliographystyle{unsrt}

\end{document}